\documentclass[twocolumn,showpacs,preprintnumbers,amsmath,amssymb,floatfix,superscriptaddress]{revtex4-1}
\usepackage{graphicx}
\usepackage{amsmath,amsfonts,amssymb}
\usepackage{graphicx}
\usepackage{color}

\usepackage[normalem]{ulem}
\usepackage{todonotes}
\usepackage[bookmarks, 
            breaklinks, 
            pdfstartview=FitH, 
            pdftitle={},
            pdfauthor={},
            pdfkeywords={}
            ]{hyperref}
\usepackage{ulem}

\def\beq{\begin{equation}}
\def\eeq{\end{equation}}

\newcommand{\spl}[1]{\begin{align}\begin{split} #1 \end{split} \end{align}}

\newcommand{\vvec}[2]{\begin{pmatrix} #1 \\ #2 \end{pmatrix}}

\newcommand{\matrnop}[4]{\begin{matrix} #1 & #2 \\ #3 &  #4 \end{matrix}}

\newcommand{\affA}{QUANTUM, Institut f{\"u}r Physik, Johannes Gutenberg-Universit{\"a}t Mainz, D-55099 Mainz, Germany}

\newcommand{\affD}{Zentrum f\"ur Optische Quantentechnologien and The Hamburg Centre for Ultrafast Imaging, Universit\"at Hamburg, Luruper Chaussee 149, D-22761 Hamburg}
\newcommand{\affE}{Faculty of Physics, University of Warsaw, PL-00-681 Warsaw, Poland}
\newcommand{\affF}{Center for Integrated Quantum Science and Technology, Institute for Complex Quantum Systems, University of Ulm, Albert-Einstein-Allee 11, D-89069 Ulm, Germany}

\begin{document}

\title{Generalised Kronig-Penney model for ultracold atomic quantum systems}

\author{A.~Negretti}\affiliation{\affD}
\author{R.~Gerritsma}\affiliation{\affA}
\author{Z.~Idziaszek}\affiliation{\affE}
\author{F.~Schmidt-Kaler}\affiliation{\affA}
\author{T.~Calarco}\affiliation{\affF}

\begin{abstract}
We study the properties of a quantum particle interacting with a one dimensional structure of equidistant scattering centres. 
We derive an analytical expression for the dispersion relation and for the Bloch functions in the presence of 
both even and odd scattering waves within the pseudopotential approximation. 
This generalises the well-known solid-state physics text-book result known as the Kronig-Penney model. Our generalised model can be 
used to describe systems such as degenerate Fermi gases interacting with ions or with another neutral atomic species confined in an optical lattice, thus enabling 
the investigation of polaron or Kondo physics within a simple formalism. We focus our attention on the specific atom-ion system and compare our findings with quantum 
defect theory. Excellent agreement is obtained within the regime of validity of the pseudopotential approximation. This enables us to derive a Bose-Hubbard Hamiltonian 
for a degenerate quantum Bose gas in a linear chain of ions. 
\end{abstract}

\date{\today}

\maketitle

%
%

\section{Introduction}

The Kronig-Penney (KP) model is an analytically solvable model of a one-dimensional (1D) crystalline solid in which the electron-nuclei interactions are replaced by contact potentials of the Dirac-delta form~\cite{Kronig1931}. It is often used in text-books of solid-state physics to help clarify the emergence of an electronic band structure~\cite{Kittel2005}. Although mostly considered to be of academic and educational - rather than quantitative - interest, a number of laboratory systems have become available for which the KP-model provides a useful starting point. These experiments range from solid state and surface science where one-dimensional structures of atoms can be surface-deposited in scanning tunnelling microscopy~\cite{Nilius2002,Ortega2006,Oncel2008} to cold atomic systems in periodic potentials~\cite{Bloch2005}. In such model systems, a setup in which one species forms a lattice of atoms - or ions - through which a second, untrapped, species can move, would create a situation reminiscent of the KP-model. This is particularly so, since the interaction between the lattice of atoms - or ions - and the untrapped atoms could be accurately described by Fermi's zero-range pseudopotential~\cite{Fermi1936,Huang1987} that is commonly employed in ultracold atomic physics. The considered system may be of significant interest in constructing a quantum simulator of crystalline solids~\cite{Cirac2012,Bloch2012,Blatt2012}. In particular, the inherent quantum nature of the lattice potential, that may have spin or phononic degrees of freedom, sets the system apart from quantum simulators based exclusively on (classical) optical lattice potentials~\cite{Bissbort2013}. In this way, the structure of a natural solid is more accurately emulated, making it possible to simulate electron-phonon coupling or Kondo physics. 

\begin{figure}[h!]
\includegraphics*[scale=0.18]{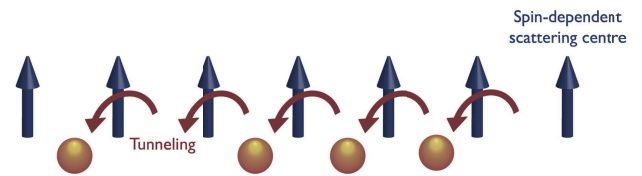}
\caption[]{%
(Color online). We consider a 1D system of scattering centers with spin degrees of freedom through which particles can move.
}
\label{fig:setup}
\end{figure}

In this paper, we generalise the KP-model, where a series of equidistant Dirac's deltas forms the periodic potential as shown in Fig.~\ref{fig:setup}, by including the first derivative of that delta potential. This allows our model to be used in the physical relevant scenario in which both $s$-wave and $p$-wave scattering are present. As in the original Kronig-Penney model, our model results in easy-to-interpret analytical expressions. We compare the model to a full numerical calculation based on quantum defect theory (QDT) for the case in which the crystal is formed by ions and the untrapped species is formed by neutral atoms. Excellent agreement is achieved within the validity regime of the pseudopotential approximation. We derive the corresponding Bloch and Wannier functions and demonstrate that the low-energy limit of the system is described by a Bose-Hubbard (BH) Hamiltonian, where the coupling term $J$ depends on the relative spin-orientation of the atom and ion~\cite{Gerritsma2012,Joger2014}. Recent advances in experiments combining atomic and ionic species may put the hybrid atom-ion system within experimental reach~\cite{Haerter2014}, and our model is of relevance to approaches where the ions are replaced by a second neutral species as well~\cite{Micheli2006, Bruderer2007,Lan2013}.

The paper is organised as follows: In Sec.~\ref{sec:pseudo} we briefly review the derivation of the pseudopotentials for even- and odd-waves as discussed 
in Ref.~\cite{Girardeau2004}. In section~\ref{sec:gKP} we present our generalised KP-model, while in Sec.~\ref{sec:qdt} we compare it to a quantum defect theory 
calculation for an atom in an ion chain. Thereafter, in Sec.~\ref{sec:bloch} we obtain the corresponding Bloch and Wannier functions and in Sec.~\ref{sec:application} 
we apply the model to the hybrid atom-ion system in order to derive the BH Hamiltonian. We discuss our findings and outlooks in Sec.~\ref{sec:conc}. Finally, 
in the Appendix~\ref{sec:appB} we provide details on the 1D asymptotic solutions of QDT for the hybrid atom-ion system, in the Appendix~\ref{sec:appA} on 
the derivation of the energy-dependent scattering lengths, and in the Appendix~\ref{sec:appC} on the computation of the Wannier-Kohn wavefunctions. 

%
%

\section{Pseudopotentials for even- and odd-waves}
\label{sec:pseudo}

The so-called zero-range pseudopotential~\cite{Fermi1936,Huang1987}, first introduced by Enrico Fermi in order to describe the short-range interaction of quasi free 
Rydberg electrons~\cite{Fermi1936}, has proven to be very successful to describe the interaction between ultracold atoms~\cite{Castin2001a,Pitaevskii2003}. The 
peculiarity of this potential is that it relies on a single parameter, the so-called 3D $s$-wave scattering length, which enables the description of the scattering at 
distances larger than the effective range of the van der Waals interaction with very good accuracy. 

Under particular circumstances, it is also possible to describe the collision between ultracold bosonic atoms in a waveguide, that is, in a 1D setting where a tight transverse 
confinement is given, by a pseudopotential~\cite{Olshanii1998}. Let us remind how this works: Assuming no trapping potential for the two colliding atoms along the 
longitudinal direction $x$, while transversally they experience a strong harmonic confinement, the asymptotic form of the wave function in relative coordinates reads~\cite{Olshanii1998}:

\begin{align}
\label{eq:asympexp}
\lim_{\vert x\vert \rightarrow \infty} \!\!\Psi(x,\rho) = \left\{ e^{ik_x x} + f_e e^{ik_x \vert x\vert}  +  \text{sign}(x) f_o e^{ik_x \vert x\vert} \right\}\phi_g(\rho),
\end{align}
where $\phi_g(\rho)$ is the ground state of the transverse harmonic trap and sign($x$) is the sign function. Here the first term in the curly brackets represents 
the incident wave, the second and third terms give the even and odd scattered waves, respectively. Besides, $f_e$ and $f_o$ are the 1D scattering amplitudes for
the even- and odd-waves, respectively. In Ref.~\cite{Olshanii1998} it has been shown that while $f_e\ne 0$, the odd-wave scattering amplitude $f_o$ vanishes. 
This is a consequence of the bosonic symmetry of the colliding atoms. Furthermore, the odd-wave scattering amplitude vanishes also for a 1D delta 
pseudopotential~\cite{Olshanii1998}. 

Now if we have a spin-$\frac{1}{2}$-polarized atomic Fermi gas, depending on the internal state of the fermions and because the total fermionic wave function 
must be antisymmetric, both even-wave and odd-wave scattering can occur. The same happens if we have distinguishable particles like an atom and an ion. 
In these cases, this implies that the 1D delta potential is not sufficient to describe the interaction of two fermions in a symmetric spin state or of two distinguishable 
particles. This is clear from Eq.~(\ref{eq:asympexp}) as in these cases we have to expand the general solution to the Schr\"odinger equation both on even and odd functions. 

Almost a decade ago, however, Girardeau and Olshanii have derived the analytical expressions for the even-wave and odd-wave pseudopotentials for two interacting 
fermions~\cite{Girardeau2004}. We note, however, that they can be applied to distinguishable particle too. The actual potential in this scenario is given by 
$\upsilon(x) = \upsilon_{\mathrm{1D}}^e(x) + \upsilon_{\mathrm{1D}}^o(x)$, where the even and odd (two-body) pseudopotentials are defined as~\cite{Girardeau2004}:

\begin{align}
\label{eq:pseudo-eo}
\upsilon_{\mathrm{1D}}^e(x) = g_{\mathrm{1D}}^e \hat\delta_{\pm}(x),\qquad
\upsilon_{\mathrm{1D}}^o(x) = g_{\mathrm{1D}}^o \delta^{\prime}(x)\hat\partial_{\pm}.
\end{align}
Here $g_{\mathrm{1D}}^e = -\hbar^2/\mu a_{\mathrm{1D}}^e$ and $g_{\mathrm{1D}}^o = -\hbar^2 a_{\mathrm{1D}}^o/\mu$ with $\mu$ being the relative mass 
of the fermionic - or distinguishable - particles, and $a_{\mathrm{1D}}^{e,o}$ are the 1D scattering lengths for even- and odd-waves, respectively. The apex $^{\prime}$ 
denotes the spatial derivative. Besides, the action of the two operators appearing in Eq.~(\ref{eq:pseudo-eo}) on a wave function $\psi(x)$ is given by:

\begin{align}
2\, \hat\delta_{\pm}(x)\psi(x) &= [\psi(0^+) + \psi(0^-)] \delta(x),\nonumber\\
2\, \hat\partial_{\pm}\psi(x) &= [\psi^{\prime}(0^+) + \psi^{\prime}(0^-)] ,
\end{align}
where $\psi(0^{\pm}) = \lim_{x\rightarrow 0^{\pm}}\psi(x)$. We note that the derivative of the delta potential appearing in $\upsilon_{\mathrm{1D}}^o(x)$ is a direct consequence 
of the fact that the odd part of the wavefunction in Eq.~(\ref{eq:asympexp}) is not continuous in $x=0$. 

Given this, we have all ingredients to solve the problem for a periodic potential of even- and odd-wave interactions. 

%
%

\section{Generalisation of the Kronig-Penney model}
\label{sec:gKP}

\begin{figure}
\includegraphics*[width=\columnwidth]{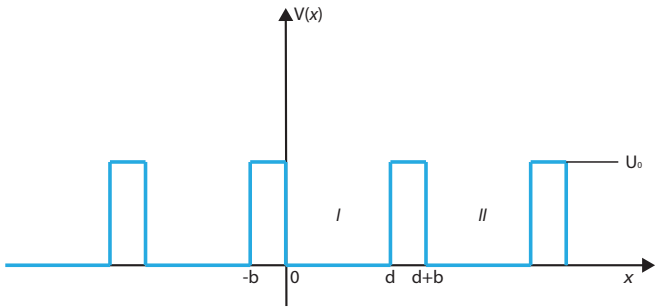}
\caption[]{%
(Color online). Sketch of the periodic potential used in the original Kronig-Penney model.
}
\label{fig:sketch}
\end{figure}

The Kronig-Penney model describes a single particle moving in a one dimensional periodic potential of rectangular 
barriers of height $U_0$ and width $b$ separated by a distance $d$ (see Fig.~\ref{fig:sketch}). 
A special relevant case is the one when the limits $U_0\rightarrow\infty$ and $b\rightarrow 0$ 
are taken, namely when the rectangular barriers are replaced by a sequence of delta potentials. 
It is this particular scenario we are interested in.

Now, from a quantum mechanical scattering point of view, the periodic potential 

\begin{align}
V_e(x) =  \sum_k \upsilon_{\mathrm{1D}}^e(x-x_k)
\end{align}
corresponds to the situation for which the cores and the moving particle are bosons, and therefore only even-wave scattering occurs~\cite{Olshanii1998}. 
Our goal here is to solve the Schr\"odinger equation for the following periodic potential 

\begin{align}
\label{eq:Vx}
V(x) =  \sum_k \upsilon_{\mathrm{1D}}^e(x - x_k) +  \sum_k \upsilon_{\mathrm{1D}}^o(x - x_k).
\end{align}

%
%

\subsection{Solution for the odd-waves}

To begin with, let us consider the case for which $\upsilon_{\mathrm{1D}}^e\equiv 0$.  Thus, we aim at the determination of the dispersion relation for the single particle 
Hamiltonian~\footnote{We note that from now on we replace the relative mass $\mu$ with the mass $m$ of the moving particle. Although the scattering process is described 
in the relative coordinates, we are now simply interested in the dynamics of the moving particle. Thus, the scattering centres are treated as fixed points in space providing only 
an external potential for the quantum particle.}

\begin{align}
\hat H_o=  -\frac{\hbar^2}{2 m}\frac{\partial^2}{\partial x^2} +  \sum_j \upsilon_{\mathrm{1D}}^o(x - x_j)
\end{align}
with $x_j = j d$ and $j\in \mathbb{Z}$. Here $m$ denotes the mass of the moving particle. To this end, we shall exploit two conditions: firstly, the behaviour of the fermionic quantum system at low energies ($k\rightarrow 0$), 
for which the wave function must satisfy the identity~\cite{Girardeau2004}

\begin{align}
\label{eq:condo}
\psi(0^+) - \psi(0^-) = - a_{\mathrm{1D}}^o [ \psi^{\prime}(0^+) + \psi^{\prime}(0^-) ].
\end{align}
This identity implies that the wave function is discontinuous in the origin, namely at the contact point, as we already pointed out at the end of Sec.~\ref{sec:pseudo}. 
This condition is different from the one of $\upsilon_{\mathrm{1D}}^e(x)$, that is,  

\begin{align}
\label{eq:conde}
\psi^{\prime}(0^+) - \psi^{\prime}(0^-) = - (a_{\mathrm{1D}}^e)^{-1} [ \psi(0^+) + \psi(0^-) ]
\end{align}
for which the derivative of the wave function has a discontinuity at the contact point. While for the latter the wave function is assumed to be continuous, in the former case 
(odd-wave) the left and right limits of the first spatial derivative are equal, namely $\psi^{\prime}(0^+) = \psi^{\prime}(0^-)$, which is a 
consequence of the antisymmetry of the wavefunction (see also Ref.~\cite{Girardeau2003}). 
We note, however, that this does not mean that the wavefunction is a continuous function.

The second condition we shall apply to solve the Schr\"odinger equation is due to the Bloch theorem~\cite{Kittel2005} which states that the wave function in the nearby interval 
($II$) is given by $\psi_{II}(x) = e^{iqd}\psi_I(x-d)$~\footnote{We note that with respect to Fig.~\ref{fig:sketch} we have set $b\rightarrow 0$ and $U_0\rightarrow \infty$.} 
(see also Fig.~\ref{fig:sketch}). Here $q$ is the Bloch vector or quasi-momentum. Putting together the two conditions we have:

\begin{align}
\label{eq:condobloch}
\psi_{II}(d^+) - \psi_I(d^-) &= - 2 a_{\mathrm{1D}}^o \psi^{\prime}_I(d^-), \nonumber\\
\psi^{\prime}_{II}(d^+) & = \psi^{\prime}_I(d^-) = e^{iqd} \psi^{\prime}_I(0^-).
\end{align}
Now we make the following Ansatz for the wave function in the interval $[0,d]$: 

\begin{align}
\label{eq:ansatzpsi}
\psi_I(x) = A \cos(k x) + B \sin(k x),
\end{align}
where $k=\sqrt{2 m E/\hbar^2}$. 
This wave function clearly solves the Schr\"odinger equation for a free particle in the interval $[0,d]$. Now by replacing~(\ref{eq:ansatzpsi}) in Eq.~(\ref{eq:condobloch}) we 
obtain a matrix equation, $\mathbf{M}\,\mathbf{C}=\mathbf{0}$, where $\mathbf{C}=(A,B)^T$ and $\mathbf{M}$ is a $2\times 2$ matrix defined as:

\begin{widetext}
\begin{align}
\mathbf{M} = \left(
\begin{array}{ccc}
\cos(k d) + 2 a_{\mathrm{1D}}^o k \sin(k d) - e^{iqd} & & \sin(k d) - 2 a_{\mathrm{1D}}^o k \cos(k d) \\
 & & \\
k \sin(k d) & & k(e^{iqd}-\cos(k d))
\end{array}
\right).
\end{align}
\end{widetext}
By imposing the condition det($\mathbf{M}$) = 0 we finally obtain the following dispersion relation: 

\begin{align}
\label{eq:dispo}
\cos(q d) = \cos(k d) + a_{\mathrm{1D}}^o k \sin(k d).
\end{align}
For the sake of completeness we provide here also the dispersion relation for the even-wave described by $\upsilon_{\mathrm{1D}}^e(x)$:

\begin{align}
\label{eq:dispe}
\cos(q d) = \cos(k d) - \frac{\sin(k d)}{a_{\mathrm{1D}}^e k}.
\end{align}
We see that for odd-waves, when $a_{\mathrm{1D}}^o \rightarrow 0$, we recover the (parabolic) spectrum of a free particle. The same 
occurs for even-waves, but when $a_{\mathrm{1D}}^e \rightarrow \infty$, which implies that $g_{\mathrm{1D}}^e \rightarrow 0$. 

\begin{figure}
\includegraphics*[width=\columnwidth]{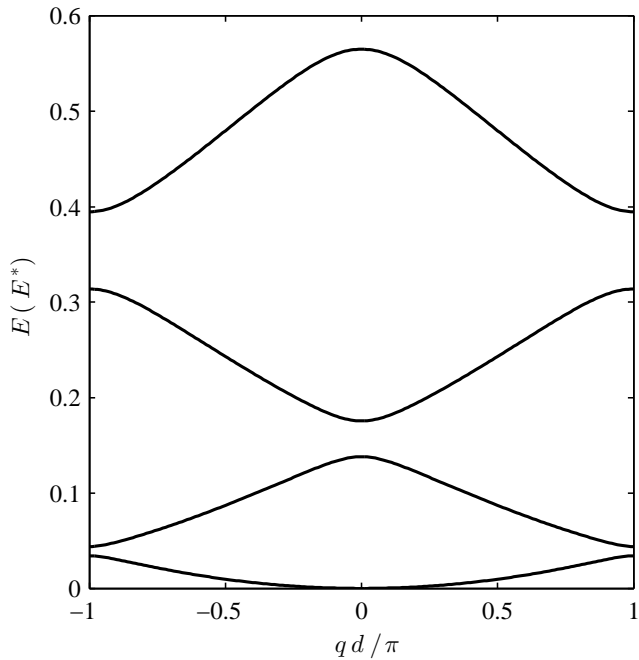}
\caption[]{%
Example of energy bands for the odd-wave dispersion relation~(\ref{eq:dispo}). 
For this band structure calculation, the parameters $d=15 R^*$, $a_{\mathrm{1D}}^e \rightarrow\infty$, and $a_{\mathrm{1D}}^o = - (\mu/m)^{1/2}R^* = - (\mu/m)^{1/2} d/15$ 
have been chosen. Note that energy and length are scaled according to the typical atom-ion interaction scales (see Sec.~\ref{sec:qdt} and in particular 
Eq.~(\ref{eq:a1d}) for details). In this unit system $\mu$ denotes the reduced mass, whereas $m$ the atom mass.
}
\label{fig:odd-exp}
\end{figure}

In Fig.~\ref{fig:odd-exp} we show a typical band structure calculation for the case of odd-wave interactions. For convenience we express energy and lengths in 
atom-ion units (see Sec.~\ref{sec:qdt}), as later we are going to discuss this system in some detail. 

We note that the size of the energy gaps grows as the energy becomes large, like in the usual KP-model. This is different with respect to the case, for instance, of an 
optical lattice with finite amplitude, where the difference between energy levels becomes smaller and smaller as the energy of the band increases. We explain this by 
noting that the behaviour is very similar to the eigenenergies of a particle in a box with perfectly rigid walls which scale as $E_n\propto n^2$ ($n$ is 
the quantum number). For a particle in such a box potential the energy difference between the levels increases. Hence, when the separation between the lattice sites 
is rather large (i.e., the pseudopotential is applicable) and because at the lattice sites the potential is infinitely large, the band structure of a particle in such a periodic 
potential is strictly connected to the energy spectrum of a particle in a box with perfectly rigid walls. 

%
%

\subsection{Solution for both even- and odd-waves}

Now we solve the general problem for which the periodic potential is given by Eq.~(\ref{eq:Vx}). To this aim, 
we apply once again the Bloch theorem and impose the conditions~(\ref{eq:condo}) and~(\ref{eq:conde}). 
This turns out to be equivalent to the following conditions: 

\begin{align}
\label{eq:condoe}
a_{\mathrm{1D}}^e\left[\psi^{\prime}_I(0^+) - \psi^{\prime}_I(d^-) e^{-iqd}\right] + \psi_I(0^+) + e^{-iqd}  \psi_I(d^-) = 0, \nonumber\\
a_{\mathrm{1D}}^o\left[\psi^{\prime}_I(0^+) + \psi^{\prime}_I(d^-) e^{-iqd}\right] + \psi_I(0^+) - e^{-iqd}  \psi_I(d^-) = 0.
\end{align}
As previously described, we use the Ansatz~(\ref{eq:ansatzpsi}) for the wavefunction in the region $I$ (see also Fig.~\ref{fig:sketch}) 
and compute the determinant of the new matrix $\mathbf{M}$, from which we obtain the new dispersion relation

\begin{align}
\label{eq:dispoe}
\cos(q d) &= \frac{(a_{\mathrm{1D}}^e + a_{\mathrm{1D}}^o) \cos(k d)}{a_{\mathrm{1D}}^e - a_{\mathrm{1D}}^o}  \nonumber\\
\phantom{=}&+ \left(k^2 a_{\mathrm{1D}}^e a_{\mathrm{1D}}^o -1\right)\frac{\sin(k d)}{(a_{\mathrm{1D}}^e - a_{\mathrm{1D}}^o) k}.
\end{align}
In the limit $a_{\mathrm{1D}}^o \rightarrow 0$ we recover~(\ref{eq:dispe}), whereas in the limit $a_{\mathrm{1D}}^e \rightarrow \infty$ we recover~(\ref{eq:dispo}). 
In the former case, this corresponds to the usual KP-model~\cite{Kittel2005} and it can be applied to non-dipolar neutral-atom 
systems where the $p$-wave interaction is negligible. On the other hand, as we will see later, for atom-ion systems an admixture of $s$-wave and $p$-wave interactions 
is possible, and therefore the correct dispersion relation becomes Eq.~(\ref{eq:dispoe}), when the pseudopotential approximation~(\ref{eq:pseudo-eo}) can be applied.

\begin{figure}
\includegraphics*[width=\columnwidth]{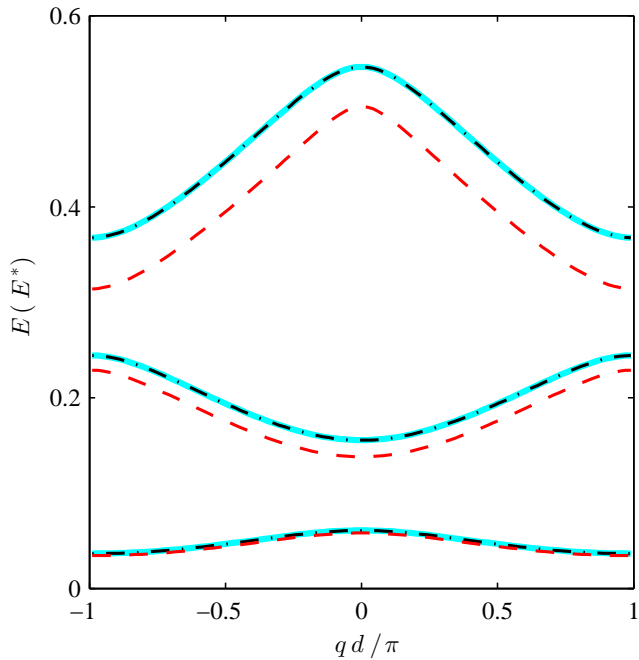}
\caption[]{%
(Color online). Energy bands for even- and odd-wave interactions. The cyan (solid thick) lines represent the result of the QDT calculation (see text), the red 
(dashed) lines are the energy bands of the generalised KP-model with energy-independent scattering lengths, while the black (dash-dot) lines with energy-dependent 
scattering lengths. The short-range phases have been chosen as in Ref.~\cite{Bissbort2013}, namely $\phi_o = -\phi_e = \pi/4$, and the ions 
are separated by $d= 15 R^*$. The corresponding energy-independent scattering lengths are $a_{\mathrm{1D}}^e = - a_{\mathrm{1D}}^o = (\mu/m)^{1/2} R^* = (\mu/m)^{1/2} d/15$. 
Note that energy and length are scaled according to the typical atom-ion interaction scales (see Sec.~\ref{sec:qdt} and in particular Eq.~(\ref{eq:a1d}) for details). 
In this unit system $\mu$ denotes the reduced mass, whereas $m$ the atom mass.
}
\label{fig:eipp-eo}
\end{figure}

In Fig.~\ref{fig:eipp-eo} we show a typical band structure calculation (red dashed lines), for which we assumed energy-independent scattering lengths (see also 
Sec.~\ref{sec:qdt}). 

%
%

\section{Comparison with quantum defect theory}
\label{sec:qdt}

In this section we compare the solutions obtained for the even and odd pseudopotentials with the solutions obtained via numerical integration of the Schr\"odinger 
equation for the atom-ion interaction potential. This scenario would correspond to an ion chain with separation $d$ between the ions, where an ultracold atom is 
free to move within the lattice, as in our recent proposal for an atom-ion quantum simulator~\cite{Bissbort2013}. 

In such a system the interaction between the moving particle and the scattering centre is caused by the electric field generated by the charge of the ion and the induced dipole 
of the atom, which at large distances has the following form in 3D: 
 
\begin{equation}
\label{eqPot}
\lim_{r\rightarrow \infty} V_{ia}(r)=-\frac{C_4}{r^4}.
\end{equation}
For $r\rightarrow 0$, Eq.~(\ref{eqPot}) does not hold 
anymore and the potential becomes strongly repulsive. The exact form of the potential in this regime depends on the electronic sub-structure of the atom and ion. 
This short-range dependence can be parametrised in the ultracold regime by a single parameter, the so-called short-range phase $\phi$~\cite{Idziaszek2007}. 
The natural energy and length scales for the atom-ion interaction~(\ref{eqPot}) are $E^*=\hbar^4/(4 m^2 C_4)$ and $R^*=\sqrt{2 m C_4/\hbar^2}$, 
respectively. Here $m$ is the mass of the atom, $C_4=e^2\alpha_p/2$ with $e$ the charge of the ion, and $\alpha_p$ the static atomic polarisability~\footnote{Note 
that the common definition of $E^*$ and $R^*$ makes use of the reduced mass of the two-body system of an atom and an ion. However, our definition is better suited 
when the ion motions is neglected.}. In these units the energy of a free particle and the corresponding wave-vector are linked as: $E=k^2$.

The atom-ion interaction clearly has a three-dimensional character, but an effective 1D theory can be developed for systems that are tightly trapped in the transverse 
direction~\cite{Idziaszek2007}. Hence, we make such assumption for the atom and we write the transverse potential, in $E^*$ and $R^*$ units, as 
$V_{\perp}(\rho)=\alpha_{\perp} \rho^2$ with $\alpha_{\perp}  = (R^*/\ell_{\perp})^4$ and $\ell_{\perp}=\sqrt{\hbar/m \omega_{\perp}}$. On the other hand, in the axial 
direction, $x$, we assume that the atom experiences no external confinement. Following Ref.~\cite{Idziaszek2007}, we also introduce the length scale 
$R_{\perp}=\alpha_{\perp} ^{-1/6}$ (in the above atom-ion units), which denotes the distance at which the atomic trapping potential $V_{\perp}$ equals the atom-ion 
interaction. Then, it can be shown that for large distances between the atom and the ion, that is, $x\gg R_{\mathrm{1D}}=\max(R_{\perp},\ell_{\perp})$, the atom-ion 
interaction can be described effectively in 1D like $-C_4/x^4$~\cite{Idziaszek2007}. 

Given all this, the Schr\"odinger equation for a single atom interacting with a static ion located at $x_i=0$ in $E^*$ and $R^*$ units is given by:

\begin{equation}
\label{eq_schr_1d}
\left(\frac{\mathrm{d}^2}{\mathrm{d} x^2}+\frac{1}{x^4}+E\right)\psi (x)=0.
\end{equation}
For $x\rightarrow 0$, the energy $E$ can be neglected and the equation can be solved analytically:

\begin{eqnarray}
\label{eq_assymp}
\tilde{\psi}_e(x)&=&|x| \sin \left(\frac{1}{|x|}+\phi_e\right)\nonumber,\\
\tilde{\psi}_o(x)&=&x \sin \left(\frac{1}{|x|}+\phi_o\right),
\end{eqnarray}
where $\phi_{e,o}$ denote the even and odd short-range phases (we refer to the Appendix~\ref{sec:appB} for some more details on choice of the above outlined 
asymptotic solutions). These phases, also called quantum defect parameters, encapsulate the physical content of 
the short range for which the actual atom-ion interaction is unknown. The parameters are two because in 1D we can have solutions with even and odd parity. 
Furthermore, although in the following we will treat the two phases as independent parameters, we note that they are not fully independent, since they are 
related both to the transverse confinement (i.e., $\omega_{\perp}$) and to the so-called 3D short-range phase~\footnote{We note, however, that the 1D scattering 
lengths can be tuned either by the frequency of the transverse trap $\omega_{\perp}$ or by means of Feshbach resonances.}. We refer to Ref.~\cite{Idziaszek2007} 
for a more detailed discussion.

Now coming back to the single atom problem in a 1D ion chain, we note that within quantum defect theory an atomic energy eigenfunction is described by a superposition of 
the asymptotic solutions~(\ref{eq_assymp}) in the vicinity of each ion, up to a sufficiently small distance $R^*\gg x_0 \gg l_{\perp}$.  Outside $x_0$, the two solutions $\psi_e(x)$ 
and $\psi_o(x)$  are propagated numerically by means of a renormalised Numerov method~\cite{Johnson1977} up to the border of the real space unit cell for a fixed set of 
$(E,\phi_e,\phi_o)$ with the initial conditions chosen to match the asymptotic solutions smoothly at $x_0$. Since the even and odd solutions span the corresponding Hilbert 
space, the Bloch state can be written as

\spl{
\psi_q(x)=c_{e}^{(q)}\,{\psi}_e(x) + c_{o}^{(q)}\,{\psi}_o(x)
}
with coefficients $c_{e}^{(q)}$ and $c_{o}^{(q)}$ for the even and odd states, respectively.
Furthermore, in accordance with the Bloch theorem, each Bloch state fulfils
\spl{
\psi_q(x+d)=e^{iqd} \, \psi_q(x)
}
under spatial translation by a lattice distance $d$. Together, these conditions allow us to relate $E=E(k)$ to $q$ and thereby to determine the band structure, in exact analogy 
to the Kronig-Penney model solution. 

Now, if $E$ does not lie within a band gap, the associated Bloch vector $q$ is defined by the constraint of both the wave function and 
its derivative being continuous at the edge of the unit cell. If the ion is chosen to lie in the middle of the unit cell at $x=0$, symmetry implies 
$\psi_e(-\frac d 2)=\psi_e(\frac d 2)$, $\psi_o(-\frac d 2)=-\psi_o(\frac d 2)$, $\psi_e'(-\frac d 2)=-\psi_e'(\frac d 2)$, $\psi_o'(-\frac d 2)=\psi_o'(\frac d 2)$ and the matching 
conditions imply the linear relations 

\spl{
\mathbf{A}(q) \vvec{c_e^{(q)}}{c_o^{(q)}} = \vvec{0}{0},
}
where
\spl{
\mathbf{A}(q) = \left( \rule{0cm}{2em} \matrnop{(1-e^{iqd}) \psi_e(\frac d 2)}{(1+e^{iqd}) \psi_o(\frac d 2)\:}{(1+e^{iqd}) \psi_e'(\frac d 2)}{(1-e^{iqd}) \psi_o'(\frac d 2)} \right).
}
We note that this equation relates the energy (which is now hidden in the wavefunctions) to the wavenumber $q$. For normalizable, non-trivial solutions of 
(${c_e^{(q)}},{c_o^{(q)}}$) to exist, the determinant of the coefficient matrix $\mathbf{A}(q)$ has to vanish. The quasi-momentum 
corresponding to the chosen energy $E$ is then efficiently determined by using, for instance, the Nelder-Mead algorithm~\cite{Nelder1967} on the determinant 
$\det(\mathbf{A}(q))$. 

An example of such a band structure calculation is illustrated in Fig.~\ref{fig:eipp-eo}, where the thick solid lines represent the energy bands for $\phi_o = -\phi_e = \pi/4$ 
and $d= 15 R^*$. In the low energy limit, the 1D scattering lengths are related to the corresponding even and odd short-range phase as~\footnote{Although in the 
present study the units $E^*$ and $R^*$ are more suitable for numerical simulations, since the ion motion is neglected, the actual atom-ion scattering process is described 
in the relative coordinate. As a consequence also the scattering lengths must be given in units of $\bar{R}^* = \sqrt{2\mu C_4/\hbar^2}= \sqrt{\mu /m}\,R^*$~\cite{Idziaszek2007}. 
This is the reason why in Eq.~(\ref{eq:a1d}) the factor $\sqrt{\mu/m}$ is appearing. However, in our numerical simulations we expressed all lengths, and therefore the scattering 
lengths, in units of $R^*$ (i.e., $\mu\mapsto m$). Thus, in order to retrieve the correct (physical) scattering lengths, in the international system of units, one has to apply Eq.~(\ref{eq:a1d}) with the corresponding short-range phases for whichever atom-ion pair.}: 

\beq
\label{eq:a1d}
a_{\mathrm{1D}}^{e,o} = -\sqrt{\frac{\mu}{m}} R^*\cot\phi_{e,o}
\eeq
with $\mu$ being the relative mass of the atom-ion system.
By replacing these definitions in the dispersion relation~(\ref{eq:dispoe}) we obtained the energy bands that are displayed in Fig.~\ref{fig:eipp-eo} with red dashed lines. As 
it is shown, the ion separation is sufficiently large such that the KP-model can reproduce very well the lowest energy band corresponding to scattering states. However, the 
agreement is worse at higher energy bands, as expected, since the relations~(\ref{eq:a1d}) do not hold anymore. The agreement, however, can be improved by considering 
energy-dependent scattering lengths, as it will be discussed in the next section. 

To give a feeling about typical numbers, we note that the situation in Fig.~\ref{fig:eipp-eo} would correspond, for instance, to $^{171}$Yb$^+$ ions separated by 5.6~$\mu$m 
interacting with a $^{87}$Rb atom. In this case $E^*/h = 411$ Hz resulting in band gaps of tens of Hertz. For the combination $^{171}$Yb$^+$/$^6$Li, the ions would be 
1.1~$\mu$m apart with $E^*/h$ = 167~kHz leading to band gaps in the 10~kHz range. 

%
%

\subsection{Energy-dependent 1D scattering lengths}
\label{sec:endepscatt}

In order to compute the energy dependent scattering lengths, we adapted the formalism developed in Ref.~\cite{Idziaszek2011} to our 1D scenario which is summarised in 
the Appendix~\ref{sec:appA}. The formalism has been applied to the previous discussed example and the result of the band structure calculation is shown in Fig.~\ref{fig:eipp-eo} 
(black dash-dot lines). The energy dependence of the scattering lengths drastically improves the result of our generalised KP-model which is now in very good agreement 
with the QDT calculation also to higher energy bands. In order to better assess and quantify the agreement between the two approaches, we introduce the following metric:

\begin{align}
\varepsilon_n:=\max_q\left\vert E^{(n)}_{\mathrm{KP}}(q) - E^{(n)}_{\mathrm{QDT}}(q)\right\vert.
\end{align}
Here $E^{(n)}_{\mathrm{KP}}(q)$ denotes the energy dispersion relation for the $n$-th band obtained with the generalised KP-model, whereas $E^{(n)}_{\mathrm{QDT}}(q)$ the one 
obtained via QDT. For the results illustrated in Fig.~\ref{fig:eipp-eo}, the error is $\varepsilon_n/E^*<1.8\times 10^{-4}$ $\forall n=1,2,3$.

We have checked the validity of this model for some pairs of short-range phases. In particular, $(\phi_e,\phi_o)=(\pi/4,-\pi/4),\,(-\pi/4,\pi/3),\,(-\pi/4,-\pi/4+\xi)$ with 
$\xi = 10^{-4}$. In the last case, we added a small correction $\xi$ in order to facilitate the 
convergence of the Numerov method. We have also checked to which extend the pseudopotential with respect to the ion separation $d$ is applicable. 
For instance, we found that for scattering states at separations close to $d=5 R^*$ and for $\phi_o = -\phi_e = \pi/4$ the error is $\varepsilon_n/E^*<0.017$ for $n=1,2$, 
while the corresponding energy gap is about 0.43 $E^*$.

%
%

\section{Bloch and Wannier functions}
\label{sec:bloch}

Now we determine the corresponding Bloch wave functions. To this end, we make the Ansatz~\cite{Wei2009}: 

\begin{align}
\label{eq:psiBloch}
\psi_q(x) = \mathcal{N}_q(k) \left\{\sin(k x) + e^{-iqd}\zeta_q(k) \sin[k (d-x)]\right\},
\end{align}
where $\mathcal{N}_q(k)$ is a normalisation constant. 
By using, for instance, condition~(\ref{eq:conde}) we obtain:

\begin{align}
\label{eq:zeta}
\zeta_q(k) = 1+\frac{a_{\mathrm{1D}}^o k(e^{2iqd}-1)}{a_{\mathrm{1D}}^o k + e^{iqd}[a_{\mathrm{1D}}^ok\cos(kd)-\sin(kd)]}.
\end{align}
Now by imposing the Born-von Karman periodic boundary conditions we obtain the following quantisation of the quasi-momentum Bloch vector: $q_j=2\pi j/(d N_L)$ with 
$n=0,\pm 1,\pm 2\, \dots, \pm (N_L-1)/2$. Here $N_L$ denotes the number of lattice sites. Thus, we normalise the Bloch wave function within a unit cell as follows

\begin{align}
\int_{0}^{d}\mathrm{d}x\,\vert\psi_q(x)\vert^2 = \frac{1}{N_L},
\end{align}
from which we finally obtain

\begin{align}
\frac{1}{\mathcal{N}_q^2(k)} &= N_L\left[\frac{d}{2}-\frac{\sin(2kd)}{4k}\right](1+\vert\zeta_q(k)\vert^2)\nonumber\\
\phantom{=}&\!\!\!\!\!\!+\vert\zeta_q(k)\vert\cos[qd-\arg(\zeta_q(k))]\left[\frac{\sin(kd)}{k}-d\cos(kd)\right]. 
\end{align}
It is easy to check that with the above outlined definitions also the condition~(\ref{eq:condo}) is fulfilled. 
Finally, accordingly to the Bloch theorem, we define the Bloch functions in another interval (i.e., cell) as:

\begin{align}
\label{eq:bloch-kp}
\psi^{(j)}_{n,q}(x) = e^{i\theta_q} e^{iq j d} \psi^{(0)}_{n,q}(x-jd) \qquad j\in\mathbb{Z},
\end{align}
where $\psi^{(0)}_{n,q}(x)\equiv \psi_q(x)$ and the index $n$ refers to the $n$-th band with $E>0$. Here we added a global phase $\theta_q$ which depends on the 
Bloch vector $q$. This phase can be chosen in such a way that the real and imaginary parts of the Bloch functions computed within QDT and the ones of the generalised 
KP-model match as much as possible for all values of $q$. More precisely, the phase $\theta_q$ has to be chosen such that the distance between the Bloch functions 
obtained with the two approaches is minimised. This task can be accomplished with any numerical search method like the Nelder-Mead algorithm~\cite{Nelder1967}. 
With the above outlined definitions, the Bloch functions over $N_L$ lattice sites form an orthonormal basis.

\begin{figure}
\includegraphics*[width=\columnwidth]{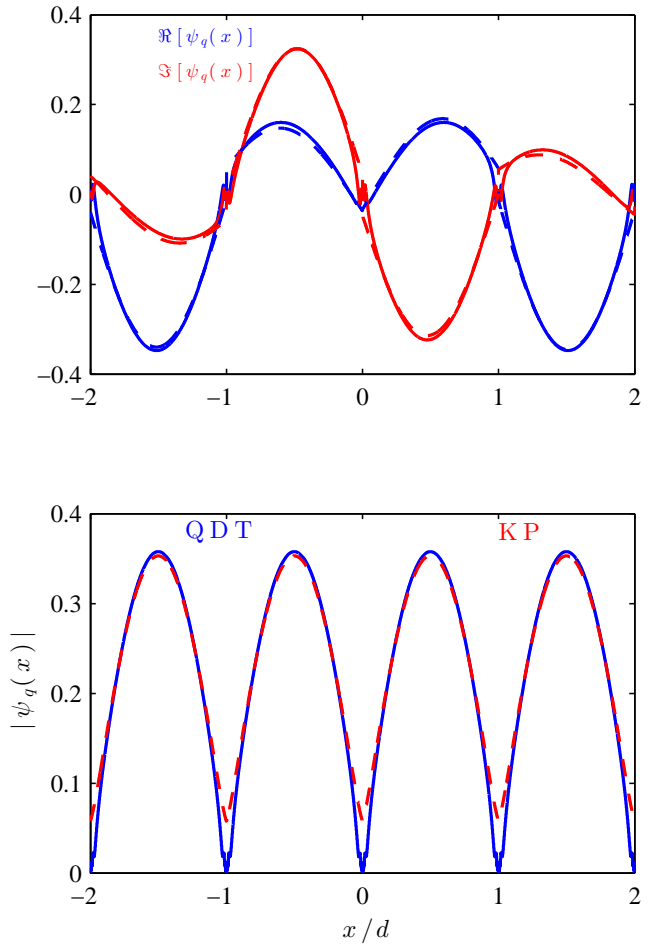}
\caption[]{%
(Color online). Upper panel: real and imaginary parts of the Bloch function corresponding to the lowest energy band of Fig.~\ref{fig:eipp-eo} for $q R^*=-0.149093$. 
Lower panel: absolute value of the Bloch wave function. In both panels the solid lines correspond to the result of QDT, whereas the dashed lines to the result of the KP-model. 
The short-range phases are $\phi_o = -\phi_e = \pi/4$ and the ions are separated by $d= 15 R^*$. For a better visualisation, we normalised the displayed Bloch 
functions on the unit cell. 
}
\label{fig:bloch}
\end{figure}

In Fig.~\ref{fig:bloch} we compare the Bloch functions obtained within QDT to the ones of our generalised KP-model for $q R^*=-0.149093$ of the lowest scattering 
energy band that is displayed in Fig.~\ref{fig:eipp-eo}. Good agreement between QDT and the generalised 
KP-model is shown, except at the scattering centres, as expected given the nature of the pseudopotential. We note that the Bloch function illustrated 
in Fig.~\ref{fig:bloch} for the generalised KP-model has been obtained with $\theta_q=0$ and that, for the sake of simplicity, hereafter we will set $\theta_q = 0$ $\forall q$. 
Despite this choice, however, the absolute values of the Bloch functions of the KP-model are in very good agreement (i.e., like in Fig.~\ref{fig:bloch}) with the ones obtained 
within QDT for all values of $q$, except at the scattering centers. On the other hand, the corresponding real and imaginary parts might not agree so well. Of course, this 
choice does not have any kind of physical relevance, but it is just a matter of numerical convenience. 

\begin{figure*}[htb!]
\centerline{
\includegraphics*[width=\columnwidth]{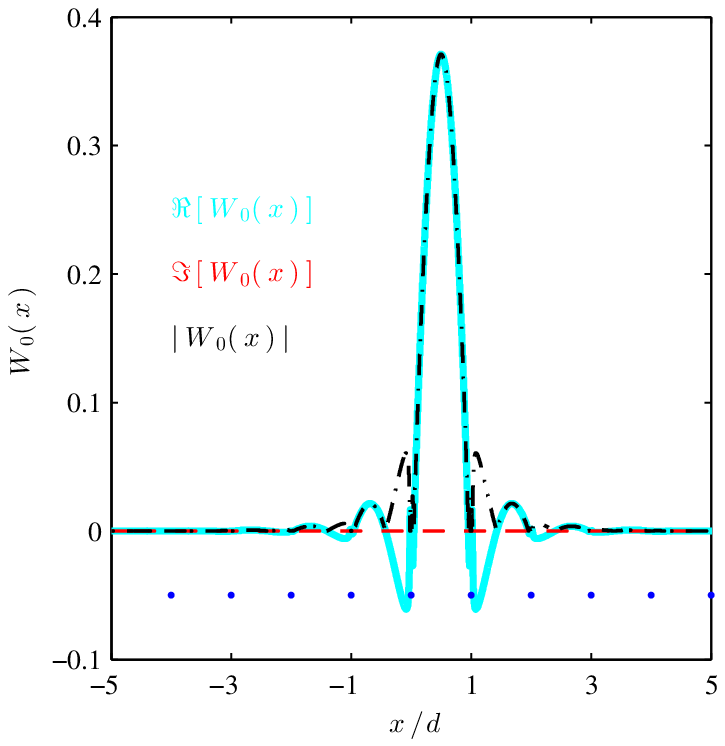}
\includegraphics*[width=\columnwidth]{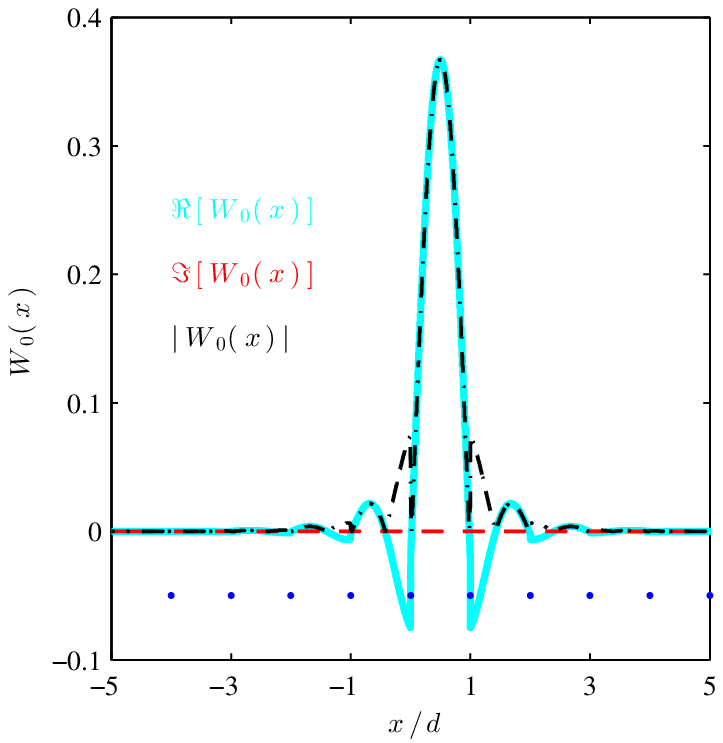}
}
\caption[]{%
(Color online). Wannier-Kohn function of the lowest scattering band within QDT (left panel) and of the generalised KP-model (right panel): the cyan (solid thick) line 
represents the real part of $W_0(x)$, the red (dashed) line its imaginary part, and the black (dash-dot) line the absolute value of the function. The blue dots serve only to 
indicate the position of the scatter sites. The short-range phases are $\phi_o = -\phi_e = \pi/4$, and the ions are separated by $d= 15 R^*$. 
}
\label{fig:wan-kohnKP-QDT}
\end{figure*}

Finally, in Fig.~\ref{fig:wan-kohnKP-QDT} we shown an example of Wannier-Kohn function computed within QDT and for our generalised KP-model. 
We note that, in order to get more localised Wannier functions, in this circumstance the Bloch functions obtained via QDT have been multiplied 
by the phase factor $e^{-iq_j d/2}$  (see also Appendix~\ref{sec:appC} for more details). We recall that the 
Wannier functions, in contrast to the Bloch functions, form a maximally localised set of orthonormal states in real space. 
Also in this case, very good agreement is obtained between the two theories, although the result of the KP-model displays 
a less smooth behaviour at the maximum and minimum of the real part of the function. This is essentially due to the singular behaviour of the pseudopotential in 
correspondence to the ion locations. 

%
%

\section{Application: The Bose-Hubbard model}
\label{sec:application}

In this section we are going to present an application of the above outlined formalism focusing again on the hybrid atom-ion system. Our goal is to 
show that a Bose-Hubbard-type Hamiltonian can be derived for an ultracold atomic ensemble of bosons immersed in an ion chain in a regime where our 
generalised KP-model can be applied (see also Fig.~\ref{fig:setup}). We will compute the hopping and interaction matrix elements and compare them 
with QDT calculations. We note, however, that the present approach can be easily generalised to other systems as well. 

\subsection{Hopping matrix elements}

In order to compute the hopping matrix elements we note that we do not need to compute the Wannier functions. There is a direct connection between 
the dispersion relation and the hopping matrix elements~\cite{Bissbort2007}:

\beq
\label{eq:hopping}
J_{l,l^{\prime}}= \frac{1}{N_L} \sum_{q_k} E_{q_k}^{(n)} e^{iq_k d(l-l^{\prime})}.
\eeq
Here again $N_L$ is the number of lattice sites~\footnote{$N_L+1$ is the number of equidistant grid points of the first Brillouin zone (BZ) $(-\pi/d,\pi/d]$.}, whereas $E_q^{(n)}$ 
denotes the corresponding $n$-th energy band as a function of the Bloch vector $q$. We note that Eq.~(\ref{eq:hopping}) can be easily obtained by expanding the the single 
particle Hamiltonian of the periodic potential on the basis of the Bloch functions and then apply the definition of Wannier function~(\ref{eq:wannnier}).

\begin{figure}
\includegraphics*[width=\columnwidth]{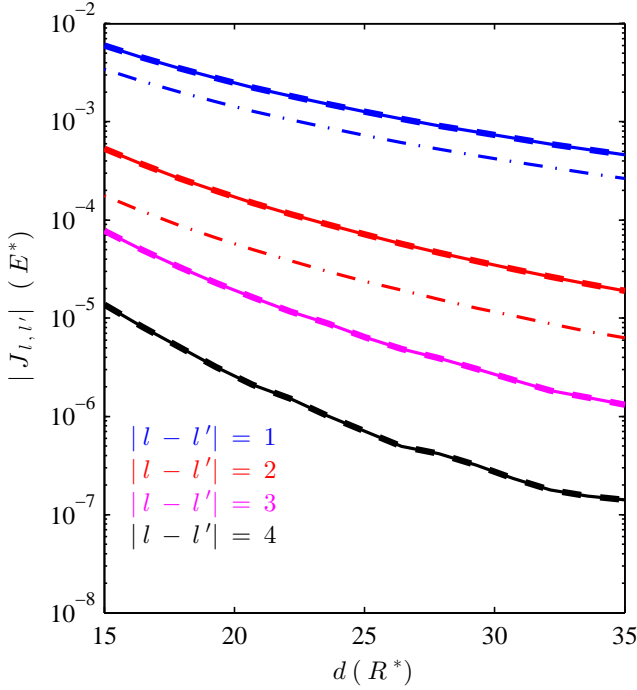}
\caption[]{%
(Color online). Hopping matrix elements vs. interionic separation $d$ for the lowest scattering band: $J_{l,l+1}$ blue line; $J_{l,l+2}$ red line; $J_{l,l+3}$ magenta line; 
$J_{l,l+4}$ black line. The thick dashed lines represent the result of QDT, whereas the thin solid lines the result of the KP-model. The short-range phases are 
$\phi_o = -\phi_e = \pi/4$ for the solid and dashed lines, whereas $\phi_o = -\phi_e = \pi/3$ for the dash-dotted lines.}
\label{fig:hoppingME}
\end{figure}

Fig.~\ref{fig:hoppingME} shows the hopping matrix elements as a function of the interionic separation $d$ for various neighbour distances $l-l^{\prime}$. As it is shown, 
for increasing separation between the ions, the strength of the hopping matrix elements decays. Additionally, the nearest neighbour matrix element is the largest one. 
Again, the calculation has been performed for the lowest scattering band. 

Finally, we also note that we have checked for $d= 15 R^*$ that, by using the definition of $J_{k,\ell}$ with the Wannier functions given in Eq.~(\ref{eq:EJU}), 
we obtain the same result. Indeed, for the pair of short-range phases $\phi_o = -\phi_e = \pi/4$, we obtain $\vert J_{k,k+1}\vert \simeq 5.8\times 10^{-3}\, E^*$ for the 
Wannier functions computed with our KP-model, while the result of QDT accordingly to Eq.~(\ref{eq:hopping}) yields $6.0\times 10^{-3}\, E^*$.

%
%

\subsection{Bose-Hubbard Hamiltonian in the single band approximation}

Let us consider an interacting atomic ensemble of ultracold bosons in the quasi 1D regime. The many-body Hamiltonian in the language of the second quantisation 
for such a system where the ions are pinned in their equilibrium positions, that is, we neglect the ion motion, can be written as 

\begin{align}
\label{eq:Hmb}
\hat H = \int\mathrm{d}x\hat\psi^{\dag}(z)\hat H_{\mathrm{sp}}\hat\psi(x)
+\frac{g_{\mathrm{1D}}}{2} \int\mathrm{d}x\hat\psi^{\dag}(x)\hat\psi^{\dag}(x)\hat\psi(x)\hat\psi(x).
\end{align}
Here $\hat H_{\mathrm{sp}} = -\frac{\hbar^2}{2 m}\frac{\partial^2}{\partial x^2} + V(x)$ is the single particle Hamiltonian, and $V(x)$ is given by Eq.~(\ref{eq:Vx}). 
Besides this, in Eq.~(\ref{eq:Hmb}) we have introduced the 1D atom-atom coupling constant $g_{\mathrm{1D}}=2\hbar\omega_{\perp}a_{aa}^s\Lambda$, where 
$\Lambda = (1- 1.4603\,a_{aa}^s/\sqrt{2}\ell_{\perp})^{-1}$ is the Olshanii correction~\cite{Olshanii1998} with $a_{aa}^s$ being the 3D $s$-wave scattering length. 
This implies that transversally the motion is frozen to the harmonic oscillator ground state. Now, by expanding the atomic quantum field operator $\hat\psi(x)$ onto 
the Wannier-Kohn basis of the lowest scattering band, that is, $\hat\psi(x) = \sum_jW_0(x-x_j)\hat b_j$ with $[\hat b_j,\hat b^{\dag}_j]=1$, the many-body 
Hamiltonian~(\ref{eq:Hmb}) becomes

\begin{align}
\label{eq:H}
\hat H=
\sum_{k,\ell}J_{k,\ell}\hat b_k^{\dag}\hat b_{\ell}
+\frac{g_{\mathrm{1D}}}{2}\sum_{k,k^{\prime},\ell,\ell^{\prime}}U_{k,k^{\prime},\ell,\ell^{\prime}}\hat b_k^{\dag}\hat b_{\ell}^{\dag}\hat b_{k^{\prime}}\hat b_{\ell^{\prime}},
\end{align}
where

\begin{align}
\label{eq:EJU}
J_{k,\ell} & = - \int\mathrm{d}x\,W^*_0(x-x_k)\hat H_{\mathrm{sp}}W_0(x-x_{\ell}),\nonumber\\
U_{k,k^{\prime},\ell,\ell^{\prime}} & = \int\mathrm{d}x\,\prod_{j=k,\ell} W^*_0(x-x_j)\prod_{s=k^{\prime},\ell^{\prime}}W_0(x-x_{s}).
\end{align}

As it has been previously shown, the relevant hopping matrix elements are $J\equiv J_{k,k+1}$. Similarly, we have checked that the most relevant interaction matrix 
elements are $U_0 \equiv U_{k,k,k,k}$, and that the elements $U_1 \equiv U_{k,k,k,k+1}$, $U_2 \equiv U_{k,k,k,k+2}$ are negligible. Indeed, for $d= 15R^*$ 
and $\phi_o = -\phi_e = \pi/4$ we obtain: $U_0^{\mathrm{QDT}} \simeq 0.1013\, E^*$, $U_0^{\mathrm{KP}} \simeq 0.0979\, E^*$, $U_1^{\mathrm{QDT}} \simeq 0.0012\, E^*$, 
$U_1^{\mathrm{KP}} \simeq 0.0013\, E^*$, $U_2^{\mathrm{QDT}} \simeq 0.0004\, E^*$, $U_2^{\mathrm{KP}} \simeq 0.0004\, E^*$, that is, the off-diagonal matrix elements are 
two orders of magnitude smaller than the onsite interact energy $U_0$, and therefore they can be safely neglected. 
This shows again the good agreement between the exact QDT calculation and the generalised KP-model. 

Finally, before we provide the final form of the BH Hamiltonian in the single band approximation, let us first check whether intraband transitions due to the atom-atom 
interaction between the two lowest scattering bands are possible. To this end, we have to compute matrix elements of this kind

\begin{align}
\label{eq:Uintrab}
U_{k,k^{\prime},\ell,\ell^{\prime}}^{\alpha_k,\alpha_{k^{\prime}},\alpha_{\ell},\alpha_{\ell^{\prime}}}  & = 
\int\mathrm{d}x\!\!\prod_{j=k,\ell} \!\!W^*_{\alpha_j}(x-x_j)\!\!\prod_{s=k^{\prime},\ell^{\prime}}\!\!W_{\alpha_s}(x-x_{s})
\end{align}
with $\alpha_k,\alpha_{k^{\prime}},\alpha_{\ell},\alpha_{\ell^{\prime}}=0,1$ being band indices. Within QDT and for $d= 15R^*$ and $\phi_o = -\phi_e = \pi/4$ we obtain: 
$U_{k,k,k,k}^{0,0,0,1}\simeq 0.0155$, $U_{k,k,k,k}^{0,0,1,1}\simeq 0.0303$, $U_{k,k,k,k}^{0,1,1,1}\simeq 0.0068$, $U_{k,k,k,k}^{1,1,1,1}\simeq 0.0376$, 
$U_{k,k,k,k+1}^{0,0,0,1}\simeq 0.0155$, and $U_{k,k,k,k+2}^{0,0,0,1}\simeq -0.0012$. We see that the most relevant matrix element is $U_{k,k,k,k}^{0,0,1,1}$. Thus, in order 
to have a small probability for intraband transitions, the following condition has to be fulfilled: 
$g_{\mathrm{1D}} n_a\ll 2 \Delta E_{01}/U_{k,k,k,k}^{0,0,1,1}$. Here $n_a$ is the atomic density in the lattice site and $ \Delta E_{01}$ denotes the bandwidth between the 
two lowest energy scattering bands. For the parameters given in Fig.~\ref{fig:eipp-eo} and for the ion-atom pair $^{171}$Yb$^+$/$^{87}$Rb we have 
$g_{\mathrm{1D}} n_a/h\ll$ 1kHz, while for the ion-atom pair $^{171}$Yb$^+$/$^{6}$Li we have $g_{\mathrm{1D}} n_a/h\ll$ 1MHz.

Given these considerations, we are left with the following Bose-Hubbard-type Hamiltonian

\begin{align}
\label{eq:BH}
\hat H_{\mathrm{BH}} = E \hat N-J\hat B+\frac{U}{2}\sum_k\hat n_k(\hat n_k -1)+\sum_k\epsilon_k\hat n_k,
\end{align}
where $\hat B = \sum_k \hat b^{\dag}_{k+1}\hat b_k+$ h.c., $\hat N = \sum_k\hat n_k = \sum_k\hat b^{\dag}_k\hat b_k$, $E\equiv J_{k,k}$, and 
$U\equiv g_{\mathrm{1D}} U_{k,k,k,k}$. 
Here, we also added the contribution of the (shallow) axial confinement of the atoms, where $\epsilon_k\approx V_T(x_k)/E^*$ 
for well localised Wannier-Kohn functions~\cite{Jaksch1998}.

Tunnelling rate $J$ and the onsite interaction energy $U$ can be controlled by the atom-ion interaction. Indeed, because the 
interaction depends on the internal state of the ion (i.e., a specific pair of short-range phases or scattering lengths is chosen), as we have shown 
in Ref.~\cite{Gerritsma2012}, we may rewrite the above BH Hamiltonian as follows:

\begin{align}
\label{eq:BHsigma}
\hat H_{\mathrm{BH}} &=-\sum_k \hat J_ k\hat b^{\dag}_{k+1}\hat b_k+\frac{1}{2}\sum_k\hat U_k\hat n_k(\hat n_k -1) \nonumber\\
\phantom{=} &+E\hat N+\sum_k\epsilon_k\hat n_k,
\end{align}
where 

\begin{align}
\label{eq:JU}
\hat{J}_k&=J_{\downarrow,k}|\downarrow_k\rangle \langle \downarrow_k|+J_{\uparrow,k}|\uparrow_k\rangle \langle \uparrow_k|,\nonumber\\
\hat{U}_k&=\sum_{\alpha,\beta=\uparrow,\downarrow}U_{\alpha,\beta;k}\vert\alpha_k\beta_{k+1}\rangle \langle \alpha_k\beta_{k+1}\vert,
\end{align}
with

\begin{align}
U_{\alpha,\beta;k} = \int\mathrm{d}x\,\prod_{\alpha=\uparrow,\downarrow} W_{\alpha}^*(x-x_k)\prod_{\beta=\uparrow,\downarrow}W_{\beta}(x-x_{k}).
\end{align}
An example of the dependence of $J_{k,k+1}$ and $J_{k,k+2}$ on the internal state of the ion is shown in Fig.~\ref{fig:hoppingME} (dash-dotted lines) 
for the pairs of short-range phases $(\phi_e,\phi_o) = (-\pi/3,\pi/3)\, \text{and}\, (-\pi/4,\pi/4)$. This shows that state-dependent tunnelling can be 
produced, and that in principle complex entangled many-body states between atoms and ions can be engineered. 

We note that, to the best of our knowledge, this is the first time that such a BH Hamiltonian is derived for the hybrid atom-ion system in an 
experimentally accessible parameter regime. Although similar Hamiltonians have been derived for other systems, especially for ultracold atoms in optical 
lattices, the one given by Eq.~(\ref{eq:BHsigma}) results from the admixture of both $s$-wave and $p$-wave interactions among the 
scattering centres, the ions, and the moving atoms. With non-dipolar neutral atoms strong $p$-wave interactions are difficult to engineer, as they
require $p$-wave Feshbach resonances. Hence, this offers additional controllability to the atom-ion system with respect 
to neutral atomic systems. In particular, the possibility to control independently the hopping and the onsite terms defined in Eq.~(\ref{eq:JU}) via 
the optical control of the internal state of the ions is an interesting alternative to neutral-atom systems, where more elaborated schemes have been 
devised to engineer such state-dependent couplings. This enables to explore the physics of lattice models and entanglement generation between 
the moving particles and the scatterers in such hybrid systems. 

Finally, we note that the Hamiltonian~(\ref{eq:BHsigma}) resembles the one of an atomic ensemble in interaction with a field cavity mode. Such an interaction 
enables cavity-mediated long-range atom-atom interactions~\cite{Maschler2005}. In our system the quantum potential is provided by the atom-ion interaction 
where the atomic back-action on the quantum lattice potential may generate atom-ion entanglement via phonons~\cite{Gerritsma2012,Bissbort2013}. Furthermore, 
we note that with our setup we can also study the analogue of a single atom transistor~\cite{Micheli2006}, where one ion of the chain can eventually suppress the 
atomic tunnelling via the state-dependent atom-ion interaction.

%
%

\section{Conclusions}
\label{sec:conc}

In this work we have presented the solution of a generalised Kronig-Penney model where both $s$-wave and $p$-wave scattering are present. We derived 
analytical expressions for the dispersion relation and the Bloch functions in this model. We have compared the results of the KP-model with a quantum defect 
theory calculation for the hybrid atom-ion system and found very good agreement between the two theories. For describing the higher energy bands, the energy 
dependence of the 1D atom-ion scattering lengths needs to be taken into account. This also enabled us to derive the low-energy Bose-Hubbard Hamiltonian 
for such a hybrid system. 

We stress that our generalised KP-model can be used to describe other systems as well, such as atoms trapped in an optical lattice or one-dimensional structures. 
For instance, in Ref.~\cite{Pupillo2008} a Bose-Hubbard model for cold atoms in a dipolar crystal has been derived, a system with features similar to the 
atom-ion system, where the usual KP-model has been used to describe the interaction between the moving atoms and the dipolar molecules. It would be interesting 
to see whether one can engineer this interaction in such a way that the $p$-wave contribution becomes important. 
Then, our generalised KP-model could be applied to make simple estimations, for instance, of the atom-phonon coupling in such lattice model or for exploring the 
underlying physics of the polaronic Fr\"ohlich Hamiltonian of a recently proposed atom-ion quantum 
simulator~\cite{Bissbort2013}. Given recent advances in experimental atomic and solid-state quantum physics, the model we analysed has not only an academic 
interest, but it is indeed applicable to systems that can be realised in current experiments, for instance, with ultracold atomic gases.

%
%

\section*{Acknowledgements}

This work was supported by the excellence cluster The Hamburg Centre for Ultrafast Imaging - Structure, Dynamics and Control of Matter at the Atomic Scale of 
the Deutsche Forschungsgemeinschaft (A.N.), by the ERC (grant 337638) (R.G.), by an internal grant of the University Mainz (Project `Hybrid atom-ion microtrap') 
(R.G.), by the EU integrated project SIQS (T.C., F.S.), by the Deutsche Forschungsgemeinschaft via the SFB/TRR21 ``Co.Co.Mat.'' (T.C.), and by the National Center for 
Science Grant No. DEC-2011/01/B/ST2/02030 (Z.I.). The authors acknowledge useful discussions with Krzysztof Jachymski and Michal Krych. A. N. acknowledges 
conversations with Panagiotis Giannakeas, Ulf Bissbort, and Johannes Schurer.

%
%

\appendix
\section{Short-range asymptotic solutions}
\label{sec:appB}

In Ref.~\cite{Idziaszek2007} a 1D QDT for ultracold atom-ion collisions has been developed. There it has been shown that 

\begin{align}
\tilde{\psi}_e(x) & = \vert x\vert \sin\left[R^*/\vert x\vert + \phi_e(k)\right] \qquad x\ll\sqrt{R/k}\nonumber\\
\tilde{\psi}_o(x) & = x \sin\left[R^*/\vert x \vert+ \phi_o(k)\right] \qquad x\ll\sqrt{R/k}\nonumber\\
\end{align}
are the even and odd solutions of the Schr\"odinger equation

\begin{align}
\label{eq:de}
\frac{\hbar^2}{2 m} \frac{\partial^2}{\partial x^2}\psi(x) + \frac{C_4}{\vert x\vert ^4}\psi(x) = 0.
\end{align}
Here we would like to show that such solutions are the most general ones of the 1D scattering problem with the $x^{-4}$ potential 
and that two independent quantum defect parameters, $\phi_{e,o}$, are indeed necessary for a full description of the scattering process. 

To this end, let us start from the following linearly independent solutions

\begin{align}
\label{eq:soljoel}
\psi_1(x) & = x \sin\left(R^*/x\right) \nonumber\\
\psi_2(x) & = x \cos\left(R^*/x\right) \nonumber\\
\end{align}
of the second order differential equation~(\ref{eq:de}). We note that their linear combination $\psi(x) = A \psi_1(x) + B \psi_2(x)$ with $A,B$ constants is 
sufficient to represent any solution of Eq.~(\ref{eq:de}) for $x>0$ (the same holds for $x<0$). Now we can construct basis for the even and the odd solutions to 
represent the bosonic and fermionic wavefunctions, respectively, that is,

\begin{align}
\label{eq:e12}
\psi_{e_1}(x) & = \vert x\vert \sin\left(R^*/\vert x \vert\right) \nonumber\\
\psi_{e_2}(x) & = \vert x\vert \cos\left(R^*/\vert x \vert\right) \nonumber\\
\end{align}
for bosons, and 

\begin{align}
\label{eq:o12}
\psi_{o_1}(x) & = x\sin\left(R^*/\vert x \vert\right) \nonumber\\
\psi_{o_2}(x) & = x \cos\left(R^*/\vert x \vert\right) \nonumber\\
\end{align}
for fermions. Finally, one can easily show that the linear combination

\begin{align}
\psi_e(x) = A _{e_1}\psi_{e_1}(x) + A_{e_1}\psi_{e_2}(x) = A_e \vert x\vert \sin(R^*/\vert x\vert + \phi_e)
\end{align}
is equivalent to $\tilde{\psi}_e(x)$, where the linear combination

\begin{align}
\psi_o(x) = A _{o_1}\psi_{o_1}(x) + A_{o_1}\psi_{o_2}(x) = A_o x \sin(R^*/\vert x\vert + \phi_o)
\end{align}
is equivalent to $\tilde{\psi}_o(x)$. Here $\phi_e = \tan(A_{e_2}/A_{e_1})$, $\phi_o = \tan(A_{o_2}/A_{o_1})$, and $A_e,\,A_o$ are some constants. 
Hence, this shows that the pair $\tilde{\psi}_{e,o}(x)$ represents the most general pair of solutions and that we actually need two independent 
quantum defect parameters for a 1D description of the atom-ion scattering process.

%
%

\section{Determination of the energy-dependent scattering lengths}
\label{sec:appA}

In order to obtain the $k$-dependence of $a_{\mathrm{1D}}^{e,o}$ we adapt the general 3D theory developed in Ref.~\cite{Idziaszek2011} for the atom-ion scattering 
process in the absence of external confinement to our pure 1D scenario. First we note that the Schr\"odinger equation~(\ref{eq_schr_1d}) is very similar to the radial 
equation of the 3D scattering problem. For the sake of completeness we provide that equation~\cite{Idziaszek2011}:

\begin{align}
\label{eq_schr_3d}
\frac{\partial^2 F(r)}{\partial r^2}+\left(
E - \frac{\ell(\ell+1)}{r^2}+\frac{1}{r^4}
\right)F(r)=0\qquad r>0.
\end{align}
Here $F(r)$ is the radial part of the 3D atomic wavefunction and $\ell$ is the partial wave quantum number. For $\ell = 0$ Eq.~(\ref{eq_schr_3d}) reduces to precisely Eq.~(\ref{eq_schr_1d}), and therefore we can apply the theory of Ref.~\cite{Idziaszek2011} by setting $\ell =0$ and by replacing the 3D short-range phase with $\phi_{e,o}$, 
whenever needed. 

For the sake of clarity, we illustrate the main steps which enables us to compute the energy-dependent scattering lengths 

\beq
a_{\mathrm{1D}}^{e,o}(k) = -\frac{\tan\delta_{e,o}(k)}{k}
\eeq
with $k\mapsto k R^* = \sqrt{E/E^*}$. 
To this end, we need to compute the energy-dependent phase shifts $\delta_{e,o}(k)$, which are defined as~\cite{Idziaszek2011}:

\begin{align}
\tan\delta_{e,o}(k) \!= \!\frac{A_{-\nu}(\phi_{e,o})m_{-\nu}\cos\eta - A_{\nu}(\phi_{e,o})m_{\nu}\sin\eta}
{A_{\nu}(\phi_{e,o})m_{\nu}\cos\eta - A_{-\nu}(\phi_{e,o})m_{-\nu}\sin\eta},
\end{align}
where $A_{\nu}(\phi_{e,o}) = \sin(\phi_{e,o} - \nu\pi/2 + \pi/4)/\sin(\pi \nu)$, $\eta = \frac{\pi}{2} (\nu - \frac{1}{2})$, 
and $m_{\nu} = (4/ k)^{\nu} S({\nu})$. Here the function $S({\nu})$ is defined as:

\begin{align}
S(\nu) = \frac{b_{\infty}^+(\nu)}{b_{\infty}^-(\nu)} \times
\frac{\Gamma\left(\frac{\nu}{2}+\frac{5}{4}\right)\Gamma\left(\frac{\nu}{2}+\frac{3}{4}\right)}
{\Gamma\left(\frac{\nu}{2}+\frac{5}{4}\right) \Gamma\left(\frac{3}{4} -\frac{\nu}{2}\right)},
\end{align}
where $\Gamma(x)$ is the Euler-function, and 

\begin{align}
b_{\infty}^{\pm}(\nu) := \lim_{n\rightarrow+\infty}b_n^{\pm}(\nu),
\end{align}
\begin{align}
b_n^{\pm}(\nu) = h^{\pm}_n(\nu)h^{\pm}_{n-1}(\nu)h^{\pm}_{n-2}(\nu)\dots h^{\pm}_1(\nu),
\end{align}
\begin{align}
\label{eq:hnpm}
h^{\pm}_n(\nu) = \frac{1}{1 - \frac{E}{[(2n + 2\pm\nu)^2-1/4][(2n\pm\nu)^2-1/4]}h^{\pm}_{n+1}}.
\end{align}
Now given the parameter $\nu$ -- we will explain immediately how to determine it -- we start by setting $h_N^{\pm}=1$, 
for some sufficiently large $N$, and we calculate $h_n^{\pm}$ by means of Eq.~(\ref{eq:hnpm}) up to $n=1$. This 
enables us to compute $b_n^{\pm}(\nu)$ and, for a large $N$, also $b_{\infty}^{\pm}(\nu)$. 

Finally, in order to compute $\nu$ we can proceed by numerically solve the following equation~\cite{Idziaszek2011}

\begin{align}
\cos(\pi\nu) = 1 - \Delta [
1-\cos(\pi\sqrt{\alpha})
],
\end{align}
where $\Delta$ is an infinite determinant (independent of $\nu$):

\begin{align}
\Delta = \left\vert
\begin{array}{ccccccc}
\ddots & \vdots & \vdots & \vdots & \vdots &\vdots & \\
\cdots & 1 & \gamma_{-2} & 0 & 0 & 0 & \cdots \\
\cdots & \gamma_{-1} & 1 & \gamma_{-1} & 0 & 0 & \cdots \\
\cdots & 0 & \gamma_0 & 1 & \gamma_0 & 0 & \cdots \\
\cdots & 0 & 0 & \gamma_1 & 1 & \gamma_1 & \cdots \\
\cdots & 0 & 0 & 0 & \gamma_2 & 1 & \cdots \\
 & \vdots & \vdots & \vdots & \vdots & \vdots & \ddots 
\end{array}
\right\vert
\end{align}
with $\gamma_n = k/(4 n^2 - \alpha)$ and $\alpha = 1/4$. As pointed out in Ref.~\cite{Idziaszek2011}, $\Delta$ converges rather quickly for relatively small matrices. 
In our numerical simulations $N\sim 50$ was already sufficient.

%
%

\section{Wannier-Kohn functions}
\label{sec:appC}

Since the Bloch vector on a finite lattice is quantised, the Wannier functions are defined as the discrete Fourier transform of 
the Bloch states within each band

\begin{align}
\label{eq:wannnier}
W_n(x-x_j) = \frac{1}{\sqrt{N_L}}\sum_{q_k}\psi_{n,q_k}(x)e^{-iq_kx_j}.
\end{align}
Here $x_j = j d$ with $j\in\mathbb{Z}$. Note that the Wannier functions obey the following orthonormality condition

\begin{align}
\int\mathrm{d}x\,W_n^*(x-x_j)W_{n^{\prime}}(x-x_{j^{\prime}}) = \delta_{n,n^{\prime}}\delta_{j,j^{\prime}}
\end{align}
which is a trivial consequence of the orthonormalisation of the Bloch functions. 

\begin{figure}
\includegraphics*[width=\columnwidth]{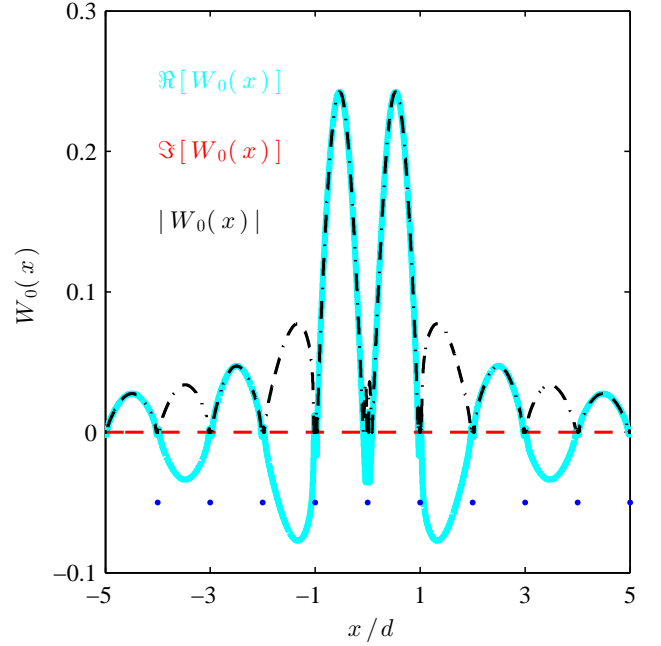}
\caption[]{%
(Color online). Wannier function of the lowest scattering band calculated within QDT: the cyan (solid thick) line represents the real part of $W_0(x)$, the red (dashed) line 
its imaginary part, and the black (dash-dot) line the modulus of the function. The blue dots serve only to indicate the position of the scatter sites. The short-range phases are 
$\phi_o = -\phi_e = \pi/4$, and the ions are separated by $d= 15 R^*$. 
}
\label{fig:wanQDT}
\end{figure}

\begin{figure}
\includegraphics*[width=\columnwidth]{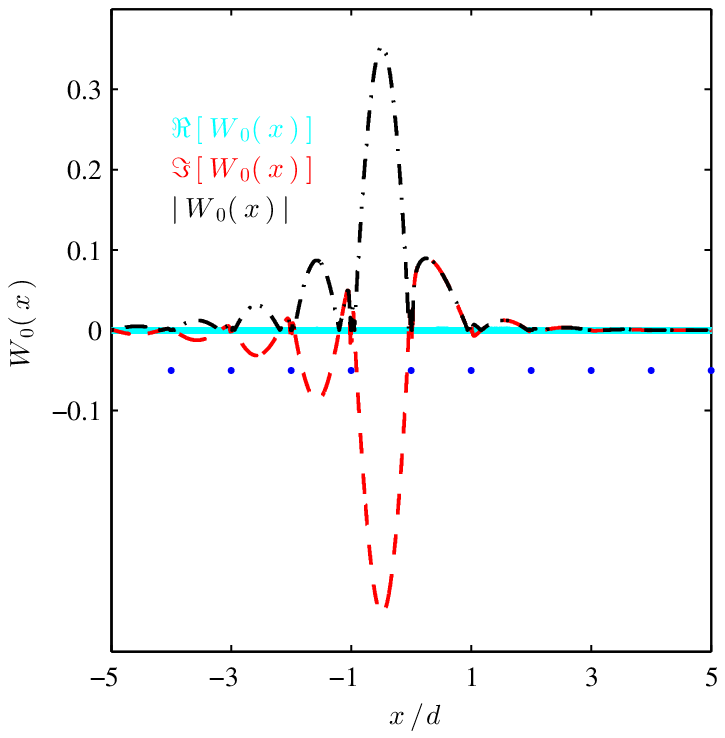}
\caption[]{%
(Color online). Wannier-Kohn function of the lowest scattering band within QDT: the cyan (solid thick) line represents the real part of $W_0(x)$, the red (dashed) line its 
imaginary part, and the black (dash-dot) line the modulus of the function. The blue dots serve only to indicate the position of the scatter sites. The short-range phases are 
$\phi_o = -\phi_e = \pi/4$, and the ions are separated by $d= 15 R^*$. 
}
\label{fig:wan-kohnQDT}
\end{figure}

In Fig.~\ref{fig:wanQDT} an instance of a Wannier function calculated within QDT is illustrated. As it is shown, the function is clearly not localised around a unit cell. 
In order to overcome this issue, we have applied Kohn's prescription~\cite{Kohn1959}, namely we multiplied the Bloch functions by a constant phase factor computed 
within QDT such that $\Re[\psi_{0,q}(0)]=0$, whereas for our generalised KP-model it has been computed such that $\Im[\psi_{0,q}(0)]=0$. 
The reason of the different factors between the two approaches is due to the fact that within QDT all Bloch functions are (in principle) zero for $x=0$, while this is not the 
case for the functions of our generalised KP-model. The result of this transformed function is shown in Fig.~\ref{fig:wan-kohnQDT} for the case of QDT, which shows a more 
localised Wannier function, even though a series of smaller peaks are observed at different lattice sites. To improve further the localisation, we have multiplied the Bloch 
functions of Bloch vector $q_j$ obtained within QDT by $e^{-i q_j d/2}$ and the result is shown in Fig.~\ref{fig:wan-kohnKP-QDT} (left panel). In this case, the localisation 
is even better.

\bibliography{liter}

\begin{thebibliography}{39}%
\makeatletter
\providecommand \@ifxundefined [1]{%
 \@ifx{#1\undefined}
}%
\providecommand \@ifnum [1]{%
 \ifnum #1\expandafter \@firstoftwo
 \else \expandafter \@secondoftwo
 \fi
}%
\providecommand \@ifx [1]{%
 \ifx #1\expandafter \@firstoftwo
 \else \expandafter \@secondoftwo
 \fi
}%
\providecommand \natexlab [1]{#1}%
\providecommand \enquote  [1]{``#1''}%
\providecommand \bibnamefont  [1]{#1}%
\providecommand \bibfnamefont [1]{#1}%
\providecommand \citenamefont [1]{#1}%
\providecommand \href@noop [0]{\@secondoftwo}%
\providecommand \href [0]{\begingroup \@sanitize@url \@href}%
\providecommand \@href[1]{\@@startlink{#1}\@@href}%
\providecommand \@@href[1]{\endgroup#1\@@endlink}%
\providecommand \@sanitize@url [0]{\catcode `\\12\catcode `\$12\catcode
  `\&12\catcode `\#12\catcode `\^12\catcode `\_12\catcode `\%12\relax}%
\providecommand \@@startlink[1]{}%
\providecommand \@@endlink[0]{}%
\providecommand \url  [0]{\begingroup\@sanitize@url \@url }%
\providecommand \@url [1]{\endgroup\@href {#1}{\urlprefix }}%
\providecommand \urlprefix  [0]{URL }%
\providecommand \Eprint [0]{\href }%
\providecommand \doibase [0]{http://dx.doi.org/}%
\providecommand \selectlanguage [0]{\@gobble}%
\providecommand \bibinfo  [0]{\@secondoftwo}%
\providecommand \bibfield  [0]{\@secondoftwo}%
\providecommand \translation [1]{[#1]}%
\providecommand \BibitemOpen [0]{}%
\providecommand \bibitemStop [0]{}%
\providecommand \bibitemNoStop [0]{.\EOS\space}%
\providecommand \EOS [0]{\spacefactor3000\relax}%
\providecommand \BibitemShut  [1]{\csname bibitem#1\endcsname}%
\let\auto@bib@innerbib\@empty
\bibitem [{\citenamefont {de~L.~Kronig}\ and\ \citenamefont
  {Penney}(1931)}]{Kronig1931}%
  \BibitemOpen
  \bibfield  {author} {\bibinfo {author} {\bibfnamefont {R.}~\bibnamefont
  {de~L.~Kronig}}\ and\ \bibinfo {author} {\bibfnamefont {W.~G.}\ \bibnamefont
  {Penney}},\ }\href@noop {} {\bibfield  {journal} {\bibinfo  {journal} {Proc.
  R. Soc. Lond. A}\ }\textbf {\bibinfo {volume} {130}},\ \bibinfo {pages} {499}
  (\bibinfo {year} {1931})}\BibitemShut {NoStop}%
\bibitem [{\citenamefont {Kittel}(2005)}]{Kittel2005}%
  \BibitemOpen
  \bibfield  {author} {\bibinfo {author} {\bibfnamefont {C.}~\bibnamefont
  {Kittel}},\ }\href@noop {} {\emph {\bibinfo {title} {Introduction to Solid
  State Physics}}},\ \bibinfo {edition} {eighth edition}\ ed.\ (\bibinfo
  {publisher} {John Wiley \& Sons, Inc., New York},\ \bibinfo {year}
  {2005})\BibitemShut {NoStop}%
\bibitem [{\citenamefont {Nilius}\ \emph {et~al.}(2002)\citenamefont {Nilius},
  \citenamefont {Wallis},\ and\ \citenamefont {Ho}}]{Nilius2002}%
  \BibitemOpen
  \bibfield  {author} {\bibinfo {author} {\bibfnamefont {N.}~\bibnamefont
  {Nilius}}, \bibinfo {author} {\bibfnamefont {T.~M.}\ \bibnamefont {Wallis}},
  \ and\ \bibinfo {author} {\bibfnamefont {W.}~\bibnamefont {Ho}},\ }\href@noop
  {} {\bibfield  {journal} {\bibinfo  {journal} {Science}\ }\textbf {\bibinfo
  {volume} {297}},\ \bibinfo {pages} {1853} (\bibinfo {year}
  {2002})}\BibitemShut {NoStop}%
\bibitem [{\citenamefont {Ortega}\ and\ \citenamefont
  {Himpsel}(2006)}]{Ortega2006}%
  \BibitemOpen
  \bibfield  {author} {\bibinfo {author} {\bibfnamefont {J.~E.}\ \bibnamefont
  {Ortega}}\ and\ \bibinfo {author} {\bibfnamefont {F.~J.}\ \bibnamefont
  {Himpsel}},\ }\href@noop {} {\bibfield  {journal} {\bibinfo  {journal} {Lect.
  Notes Phys.}\ }\textbf {\bibinfo {volume} {715}},\ \bibinfo {pages} {147}
  (\bibinfo {year} {2006})}\BibitemShut {NoStop}%
\bibitem [{\citenamefont {Oncel}(2008)}]{Oncel2008}%
  \BibitemOpen
  \bibfield  {author} {\bibinfo {author} {\bibfnamefont {N.}~\bibnamefont
  {Oncel}},\ }\href@noop {} {\bibfield  {journal} {\bibinfo  {journal} {J.
  Phys.: Condens. Matter}\ }\textbf {\bibinfo {volume} {20}},\ \bibinfo {pages}
  {393001} (\bibinfo {year} {2008})}\BibitemShut {NoStop}%
\bibitem [{\citenamefont {Bloch}(2005)}]{Bloch2005}%
  \BibitemOpen
  \bibfield  {author} {\bibinfo {author} {\bibfnamefont {I.}~\bibnamefont
  {Bloch}},\ }\href@noop {} {\bibfield  {journal} {\bibinfo  {journal} {Nature
  Physics}\ }\textbf {\bibinfo {volume} {1}},\ \bibinfo {pages} {23} (\bibinfo
  {year} {2005})}\BibitemShut {NoStop}%
\bibitem [{\citenamefont {Fermi}(1936)}]{Fermi1936}%
  \BibitemOpen
  \bibfield  {author} {\bibinfo {author} {\bibfnamefont {E.}~\bibnamefont
  {Fermi}},\ }\href@noop {} {\bibfield  {journal} {\bibinfo  {journal} {Ricerca
  Scientifica}\ }\textbf {\bibinfo {volume} {7}},\ \bibinfo {pages} {13}
  (\bibinfo {year} {1936})}\BibitemShut {NoStop}%
\bibitem [{\citenamefont {Huang}(1987)}]{Huang1987}%
  \BibitemOpen
  \bibfield  {author} {\bibinfo {author} {\bibfnamefont {K.}~\bibnamefont
  {Huang}},\ }\href@noop {} {\emph {\bibinfo {title} {Statistical
  mechanics}}},\ \bibinfo {edition} {second edition}\ ed.\ (\bibinfo
  {publisher} {John Wiley \& Sons, Inc., New York},\ \bibinfo {year}
  {1987})\BibitemShut {NoStop}%
\bibitem [{\citenamefont {Cirac}\ and\ \citenamefont
  {Zoller}(2012)}]{Cirac2012}%
  \BibitemOpen
  \bibfield  {author} {\bibinfo {author} {\bibfnamefont {J.~I.}\ \bibnamefont
  {Cirac}}\ and\ \bibinfo {author} {\bibfnamefont {P.}~\bibnamefont {Zoller}},\
  }\href@noop {} {\bibfield  {journal} {\bibinfo  {journal} {Nat. Phys.}\
  }\textbf {\bibinfo {volume} {8}},\ \bibinfo {pages} {264} (\bibinfo {year}
  {2012})}\BibitemShut {NoStop}%
\bibitem [{\citenamefont {Bloch}\ \emph {et~al.}(2012)\citenamefont {Bloch},
  \citenamefont {Dalibard},\ and\ \citenamefont {Nascimb\`ene}}]{Bloch2012}%
  \BibitemOpen
  \bibfield  {author} {\bibinfo {author} {\bibfnamefont {I.}~\bibnamefont
  {Bloch}}, \bibinfo {author} {\bibfnamefont {J.}~\bibnamefont {Dalibard}}, \
  and\ \bibinfo {author} {\bibfnamefont {S.}~\bibnamefont {Nascimb\`ene}},\
  }\href@noop {} {\bibfield  {journal} {\bibinfo  {journal} {Nat. Phys.}\
  }\textbf {\bibinfo {volume} {8}},\ \bibinfo {pages} {267} (\bibinfo {year}
  {2012})}\BibitemShut {NoStop}%
\bibitem [{\citenamefont {Blatt}\ and\ \citenamefont {Roos}(2012)}]{Blatt2012}%
  \BibitemOpen
  \bibfield  {author} {\bibinfo {author} {\bibfnamefont {R.}~\bibnamefont
  {Blatt}}\ and\ \bibinfo {author} {\bibfnamefont {C.~F.}\ \bibnamefont
  {Roos}},\ }\href@noop {} {\bibfield  {journal} {\bibinfo  {journal} {Nat.
  Phys.}\ }\textbf {\bibinfo {volume} {8}},\ \bibinfo {pages} {277} (\bibinfo
  {year} {2012})}\BibitemShut {NoStop}%
\bibitem [{\citenamefont {Bissbort}\ \emph {et~al.}(2013)\citenamefont
  {Bissbort}, \citenamefont {Cocks}, \citenamefont {Negretti}, \citenamefont
  {Idziaszek}, \citenamefont {Calarco}, \citenamefont {Schmidt-Kaler},
  \citenamefont {Hofstetter},\ and\ \citenamefont {Gerritsma}}]{Bissbort2013}%
  \BibitemOpen
  \bibfield  {author} {\bibinfo {author} {\bibfnamefont {U.}~\bibnamefont
  {Bissbort}}, \bibinfo {author} {\bibfnamefont {D.}~\bibnamefont {Cocks}},
  \bibinfo {author} {\bibfnamefont {A.}~\bibnamefont {Negretti}}, \bibinfo
  {author} {\bibfnamefont {Z.}~\bibnamefont {Idziaszek}}, \bibinfo {author}
  {\bibfnamefont {T.}~\bibnamefont {Calarco}}, \bibinfo {author} {\bibfnamefont
  {F.}~\bibnamefont {Schmidt-Kaler}}, \bibinfo {author} {\bibfnamefont
  {W.}~\bibnamefont {Hofstetter}}, \ and\ \bibinfo {author} {\bibfnamefont
  {R.}~\bibnamefont {Gerritsma}},\ }\href@noop {} {\bibfield  {journal}
  {\bibinfo  {journal} {Phys. Rev. Lett.}\ }\textbf {\bibinfo {volume} {111}},\
  \bibinfo {pages} {080501} (\bibinfo {year} {2013})}\BibitemShut {NoStop}%
\bibitem [{\citenamefont {Gerritsma}\ \emph {et~al.}(2012)\citenamefont
  {Gerritsma}, \citenamefont {Negretti}, \citenamefont {Doerk}, \citenamefont
  {Idziaszek}, \citenamefont {Calarco},\ and\ \citenamefont
  {Schmidt-Kaler}}]{Gerritsma2012}%
  \BibitemOpen
  \bibfield  {author} {\bibinfo {author} {\bibfnamefont {R.}~\bibnamefont
  {Gerritsma}}, \bibinfo {author} {\bibfnamefont {A.}~\bibnamefont {Negretti}},
  \bibinfo {author} {\bibfnamefont {H.}~\bibnamefont {Doerk}}, \bibinfo
  {author} {\bibfnamefont {Z.}~\bibnamefont {Idziaszek}}, \bibinfo {author}
  {\bibfnamefont {T.}~\bibnamefont {Calarco}}, \ and\ \bibinfo {author}
  {\bibfnamefont {F.}~\bibnamefont {Schmidt-Kaler}},\ }\href@noop {} {\bibfield
   {journal} {\bibinfo  {journal} {Phys. Rev. Lett.}\ }\textbf {\bibinfo
  {volume} {109}},\ \bibinfo {pages} {080402} (\bibinfo {year}
  {2012})}\BibitemShut {NoStop}%
\bibitem [{\citenamefont {Joger}\ \emph {et~al.}(2014)\citenamefont {Joger},
  \citenamefont {Negretti},\ and\ \citenamefont {Gerritsma}}]{Joger2014}%
  \BibitemOpen
  \bibfield  {author} {\bibinfo {author} {\bibfnamefont {J.}~\bibnamefont
  {Joger}}, \bibinfo {author} {\bibfnamefont {A.}~\bibnamefont {Negretti}}, \
  and\ \bibinfo {author} {\bibfnamefont {R.}~\bibnamefont {Gerritsma}},\
  }\href@noop {} {\bibfield  {journal} {\bibinfo  {journal} {Phys. Rev. A}\
  }\textbf {\bibinfo {volume} {89}},\ \bibinfo {pages} {063621} (\bibinfo
  {year} {2014})}\BibitemShut {NoStop}%
\bibitem [{\citenamefont {H{\"a}rter}\ and\ \citenamefont
  {Denschlag}(2014)}]{Haerter2014}%
  \BibitemOpen
  \bibfield  {author} {\bibinfo {author} {\bibfnamefont {A.}~\bibnamefont
  {H{\"a}rter}}\ and\ \bibinfo {author} {\bibfnamefont {J.~H.}\ \bibnamefont
  {Denschlag}},\ }\href@noop {} {\bibfield  {journal} {\bibinfo  {journal}
  {Contemporary Physics}\ }\textbf {\bibinfo {volume} {55}},\ \bibinfo {pages}
  {33} (\bibinfo {year} {2014})}\BibitemShut {NoStop}%
\bibitem [{\citenamefont {Micheli}\ and\ \citenamefont
  {Zoller}(2006)}]{Micheli2006}%
  \BibitemOpen
  \bibfield  {author} {\bibinfo {author} {\bibfnamefont {A.}~\bibnamefont
  {Micheli}}\ and\ \bibinfo {author} {\bibfnamefont {P.}~\bibnamefont
  {Zoller}},\ }\href@noop {} {\bibfield  {journal} {\bibinfo  {journal} {Phys.
  Rev. A}\ }\textbf {\bibinfo {volume} {73}},\ \bibinfo {pages} {043613}
  (\bibinfo {year} {2006})}\BibitemShut {NoStop}%
\bibitem [{\citenamefont {Bruderer}\ \emph {et~al.}(2007)\citenamefont
  {Bruderer}, \citenamefont {Klein}, \citenamefont {Clark},\ and\ \citenamefont
  {Jaksch}}]{Bruderer2007}%
  \BibitemOpen
  \bibfield  {author} {\bibinfo {author} {\bibfnamefont {M.}~\bibnamefont
  {Bruderer}}, \bibinfo {author} {\bibfnamefont {A.}~\bibnamefont {Klein}},
  \bibinfo {author} {\bibfnamefont {S.~R.}\ \bibnamefont {Clark}}, \ and\
  \bibinfo {author} {\bibfnamefont {D.}~\bibnamefont {Jaksch}},\ }\href@noop {}
  {\bibfield  {journal} {\bibinfo  {journal} {Phys. Rev. A}\ }\textbf {\bibinfo
  {volume} {76}},\ \bibinfo {pages} {011605} (\bibinfo {year}
  {2007})}\BibitemShut {NoStop}%
\bibitem [{\citenamefont {Lan}\ and\ \citenamefont {Lobo}(2014)}]{Lan2013}%
  \BibitemOpen
  \bibfield  {author} {\bibinfo {author} {\bibfnamefont {Z.}~\bibnamefont
  {Lan}}\ and\ \bibinfo {author} {\bibfnamefont {C.}~\bibnamefont {Lobo}},\
  }\href@noop {} {\bibfield  {journal} {\bibinfo  {journal} {Phys. Rev. A}\
  }\textbf {\bibinfo {volume} {90}},\ \bibinfo {pages} {033627} (\bibinfo
  {year} {2014})}\BibitemShut {NoStop}%
\bibitem [{\citenamefont {Girardeau}\ and\ \citenamefont
  {Olshanii}(2004)}]{Girardeau2004}%
  \BibitemOpen
  \bibfield  {author} {\bibinfo {author} {\bibfnamefont {M.~D.}\ \bibnamefont
  {Girardeau}}\ and\ \bibinfo {author} {\bibfnamefont {M.}~\bibnamefont
  {Olshanii}},\ }\href@noop {} {\bibfield  {journal} {\bibinfo  {journal}
  {Phys. Rev. A}\ }\textbf {\bibinfo {volume} {70}},\ \bibinfo {pages} {023608}
  (\bibinfo {year} {2004})}\BibitemShut {NoStop}%
\bibitem [{\citenamefont {Castin}(2001)}]{Castin2001a}%
  \BibitemOpen
  \bibfield  {author} {\bibinfo {author} {\bibfnamefont {Y.}~\bibnamefont
  {Castin}},\ }in\ \href@noop {} {\emph {\bibinfo {booktitle} {Coherent atomic
  matter waves}}},\ \bibinfo {series and number} {Lecture Notes of Les Houches
  Summer School},\ \bibinfo {editor} {edited by\ \bibinfo {editor}
  {\bibfnamefont {C.~W.}\ \bibnamefont {R.~Kaiser}}\ and\ \bibinfo {editor}
  {\bibfnamefont {F.}~\bibnamefont {David}}}\ (\bibinfo  {publisher} {EDP
  Sciences and Springer-Verlag},\ \bibinfo {year} {2001})\ pp.\ \bibinfo
  {pages} {1--136}\BibitemShut {NoStop}%
\bibitem [{\citenamefont {Pitaevskii}\ and\ \citenamefont
  {Stringari}(2003)}]{Pitaevskii2003}%
  \BibitemOpen
  \bibfield  {author} {\bibinfo {author} {\bibfnamefont {L.}~\bibnamefont
  {Pitaevskii}}\ and\ \bibinfo {author} {\bibfnamefont {S.}~\bibnamefont
  {Stringari}},\ }\href@noop {} {\emph {\bibinfo {title} {Bose-{E}instein
  condensation}}},\ \bibinfo {series} {International Series of Monographs on
  Physics}, Vol.\ \bibinfo {volume} {116}\ (\bibinfo  {publisher} {The
  Clarendon Press Oxford University Press},\ \bibinfo {address} {Oxford},\
  \bibinfo {year} {2003})\BibitemShut {NoStop}%
\bibitem [{\citenamefont {Olshanii}(1998)}]{Olshanii1998}%
  \BibitemOpen
  \bibfield  {author} {\bibinfo {author} {\bibfnamefont {M.}~\bibnamefont
  {Olshanii}},\ }\href@noop {} {\bibfield  {journal} {\bibinfo  {journal}
  {Phys. Rev. Lett.}\ }\textbf {\bibinfo {volume} {81}},\ \bibinfo {pages}
  {938} (\bibinfo {year} {1998})}\BibitemShut {NoStop}%
\bibitem [{Note1()}]{Note1}%
  \BibitemOpen
  \bibinfo {note} {We note that from now on we replace the relative mass $\mu $
  with the mass $m$ of the moving particle. Although the scattering process is
  described in the relative coordinates, we are now simply interested in the
  dynamics of the moving particle. Thus, the scattering centres are treated as
  fixed points in space providing only an external potential for the quantum
  particle.}\BibitemShut {Stop}%
\bibitem [{\citenamefont {{Girardeau}}\ and\ \citenamefont
  {{Olshanii}}(2003)}]{Girardeau2003}%
  \BibitemOpen
  \bibfield  {author} {\bibinfo {author} {\bibfnamefont {M.~D.}\ \bibnamefont
  {{Girardeau}}}\ and\ \bibinfo {author} {\bibfnamefont {M.}~\bibnamefont
  {{Olshanii}}},\ }\href@noop {} {\bibfield  {journal} {\bibinfo  {journal}
  {eprint arXiv:cond-mat/0309396}\ } (\bibinfo {year} {2003})}\BibitemShut
  {NoStop}%
\bibitem [{Note2()}]{Note2}%
  \BibitemOpen
  \bibinfo {note} {We note that with respect to Fig.~\ref {fig:sketch} we have
  set $b\rightarrow 0$ and $U_0\rightarrow \infty $.}\BibitemShut {Stop}%
\bibitem [{\citenamefont {Idziaszek}\ \emph {et~al.}(2007)\citenamefont
  {Idziaszek}, \citenamefont {Calarco},\ and\ \citenamefont
  {Zoller}}]{Idziaszek2007}%
  \BibitemOpen
  \bibfield  {author} {\bibinfo {author} {\bibfnamefont {Z.}~\bibnamefont
  {Idziaszek}}, \bibinfo {author} {\bibfnamefont {T.}~\bibnamefont {Calarco}},
  \ and\ \bibinfo {author} {\bibfnamefont {P.}~\bibnamefont {Zoller}},\
  }\href@noop {} {\bibfield  {journal} {\bibinfo  {journal} {Phys. Rev. A}\
  }\textbf {\bibinfo {volume} {76}},\ \bibinfo {pages} {033409} (\bibinfo
  {year} {2007})}\BibitemShut {NoStop}%
\bibitem [{Note3()}]{Note3}%
  \BibitemOpen
  \bibinfo {note} {Note that the common definition of $E^*$ and $R^*$ makes use
  of the reduced mass of the two-body system of an atom and an ion. However,
  our definition is better suited when the ion motions is
  neglected.}\BibitemShut {Stop}%
\bibitem [{Note4()}]{Note4}%
  \BibitemOpen
  \bibinfo {note} {We note, however, that the 1D scattering lengths can be
  tuned either by the frequency of the transverse trap $\omega _{\perp }$ or by
  means of Feshbach resonances.}\BibitemShut {Stop}%
\bibitem [{\citenamefont {Johnson}(1977)}]{Johnson1977}%
  \BibitemOpen
  \bibfield  {author} {\bibinfo {author} {\bibfnamefont {B.~R.}\ \bibnamefont
  {Johnson}},\ }\href@noop {} {\bibfield  {journal} {\bibinfo  {journal} {J.
  Chem. Phys.}\ }\textbf {\bibinfo {volume} {67}},\ \bibinfo {pages} {4086}
  (\bibinfo {year} {1977})}\BibitemShut {NoStop}%
\bibitem [{\citenamefont {Nelder}\ and\ \citenamefont
  {Mead}(1967)}]{Nelder1967}%
  \BibitemOpen
  \bibfield  {author} {\bibinfo {author} {\bibfnamefont {J.~A.}\ \bibnamefont
  {Nelder}}\ and\ \bibinfo {author} {\bibfnamefont {R.}~\bibnamefont {Mead}},\
  }\href@noop {} {\bibfield  {journal} {\bibinfo  {journal} {Comput. J.}\
  }\textbf {\bibinfo {volume} {7}},\ \bibinfo {pages} {308} (\bibinfo {year}
  {1967})}\BibitemShut {NoStop}%
\bibitem [{Note5()}]{Note5}%
  \BibitemOpen
  \bibinfo {note} {Although in the present study the units $E^*$ and $R^*$ are
  more suitable for numerical simulations, since the ion motion is neglected,
  the actual atom-ion scattering process is described in the relative
  coordinate. As a consequence also the scattering lengths must be given in
  units of $\protect \mathaccentV {bar}016{R}^* = \protect \sqrt {2\mu
  C_4/\hbar ^2}= \protect \sqrt {\mu /m}\protect \tmspace +\thinmuskip
  {.1667em}R^*$~\cite {Idziaszek2007}. This is the reason why in Eq.~(\ref
  {eq:a1d}) the factor $\protect \sqrt {\mu /m}$ is appearing. However, in our
  numerical simulations we expressed all lengths, and therefore the scattering
  lengths, in units of $R^*$ (i.e., $\mu \DOTSB \mapstochar \rightarrow m$).
  Thus, in order to retrieve the correct (physical) scattering lengths, in the
  international system of units, one has to apply Eq.~(\ref {eq:a1d}) with the
  corresponding short-range phases for whichever atom-ion pair.}\BibitemShut
  {Stop}%
\bibitem [{\citenamefont {Idziaszek}\ \emph {et~al.}(2011)\citenamefont
  {Idziaszek}, \citenamefont {Simoni}, \citenamefont {Calarco},\ and\
  \citenamefont {Julienne}}]{Idziaszek2011}%
  \BibitemOpen
  \bibfield  {author} {\bibinfo {author} {\bibfnamefont {Z.}~\bibnamefont
  {Idziaszek}}, \bibinfo {author} {\bibfnamefont {A.}~\bibnamefont {Simoni}},
  \bibinfo {author} {\bibfnamefont {T.}~\bibnamefont {Calarco}}, \ and\
  \bibinfo {author} {\bibfnamefont {P.~S.}\ \bibnamefont {Julienne}},\
  }\href@noop {} {\bibfield  {journal} {\bibinfo  {journal} {New Journal of
  Physics}\ }\textbf {\bibinfo {volume} {13}},\ \bibinfo {pages} {083005}
  (\bibinfo {year} {2011})}\BibitemShut {NoStop}%
\bibitem [{\citenamefont {Wei}\ \emph {et~al.}(2009)\citenamefont {Wei},
  \citenamefont {Gu},\ and\ \citenamefont {Lin}}]{Wei2009}%
  \BibitemOpen
  \bibfield  {author} {\bibinfo {author} {\bibfnamefont {B.-B.}\ \bibnamefont
  {Wei}}, \bibinfo {author} {\bibfnamefont {S.-J.}\ \bibnamefont {Gu}}, \ and\
  \bibinfo {author} {\bibfnamefont {H.-Q.}\ \bibnamefont {Lin}},\ }\href@noop
  {} {\bibfield  {journal} {\bibinfo  {journal} {Phys. Rev. A}\ }\textbf
  {\bibinfo {volume} {79}},\ \bibinfo {pages} {063627} (\bibinfo {year}
  {2009})}\BibitemShut {NoStop}%
\bibitem [{\citenamefont {Bissbort}(2007)}]{Bissbort2007}%
  \BibitemOpen
  \bibfield  {author} {\bibinfo {author} {\bibfnamefont {U.}~\bibnamefont
  {Bissbort}},\ }\emph {\bibinfo {title} {Stochastic Mean-Field Theory for the
  Disordered Bose-Hubbard Model}},\ \href@noop {} {\bibinfo {type} {Diploma
  thesis}},\ \bibinfo  {school} {Goethe Universit{\"a}t Frankfurt am Main
  (Germany)}, \bibinfo {address} {available at
  www.itp.uni-frankfurt.de/cms/index.php?id=116} (\bibinfo {year}
  {2007})\BibitemShut {NoStop}%
\bibitem [{Note6()}]{Note6}%
  \BibitemOpen
  \bibinfo {note} {$N_L+1$ is the number of equidistant grid points of the
  first Brillouin zone (BZ) $(-\pi /d,\pi /d]$.}\BibitemShut {Stop}%
\bibitem [{\citenamefont {Jaksch}\ \emph {et~al.}(1998)\citenamefont {Jaksch},
  \citenamefont {Bruder}, \citenamefont {Cirac}, \citenamefont {Gardiner},\
  and\ \citenamefont {Zoller}}]{Jaksch1998}%
  \BibitemOpen
  \bibfield  {author} {\bibinfo {author} {\bibfnamefont {D.}~\bibnamefont
  {Jaksch}}, \bibinfo {author} {\bibfnamefont {C.}~\bibnamefont {Bruder}},
  \bibinfo {author} {\bibfnamefont {J.~I.}\ \bibnamefont {Cirac}}, \bibinfo
  {author} {\bibfnamefont {C.~W.}\ \bibnamefont {Gardiner}}, \ and\ \bibinfo
  {author} {\bibfnamefont {P.}~\bibnamefont {Zoller}},\ }\href@noop {}
  {\bibfield  {journal} {\bibinfo  {journal} {Phys. Rev. Lett.}\ }\textbf
  {\bibinfo {volume} {81}},\ \bibinfo {pages} {3108} (\bibinfo {year}
  {1998})}\BibitemShut {NoStop}%
\bibitem [{\citenamefont {Maschler}\ and\ \citenamefont
  {Ritsch}(2005)}]{Maschler2005}%
  \BibitemOpen
  \bibfield  {author} {\bibinfo {author} {\bibfnamefont {C.}~\bibnamefont
  {Maschler}}\ and\ \bibinfo {author} {\bibfnamefont {H.}~\bibnamefont
  {Ritsch}},\ }\href@noop {} {\bibfield  {journal} {\bibinfo  {journal} {Phys.
  Rev. Lett.}\ }\textbf {\bibinfo {volume} {95}},\ \bibinfo {pages} {260401}
  (\bibinfo {year} {2005})}\BibitemShut {NoStop}%
\bibitem [{\citenamefont {Pupillo}\ \emph {et~al.}(2008)\citenamefont
  {Pupillo}, \citenamefont {Griessner}, \citenamefont {Micheli}, \citenamefont
  {Ortner}, \citenamefont {Wang},\ and\ \citenamefont {Zoller}}]{Pupillo2008}%
  \BibitemOpen
  \bibfield  {author} {\bibinfo {author} {\bibfnamefont {G.}~\bibnamefont
  {Pupillo}}, \bibinfo {author} {\bibfnamefont {A.}~\bibnamefont {Griessner}},
  \bibinfo {author} {\bibfnamefont {A.}~\bibnamefont {Micheli}}, \bibinfo
  {author} {\bibfnamefont {M.}~\bibnamefont {Ortner}}, \bibinfo {author}
  {\bibfnamefont {D.-W.}\ \bibnamefont {Wang}}, \ and\ \bibinfo {author}
  {\bibfnamefont {P.}~\bibnamefont {Zoller}},\ }\href@noop {} {\bibfield
  {journal} {\bibinfo  {journal} {Phys. Rev. Lett.}\ }\textbf {\bibinfo
  {volume} {100}},\ \bibinfo {pages} {050402} (\bibinfo {year}
  {2008})}\BibitemShut {NoStop}%
\bibitem [{\citenamefont {Kohn}(1959)}]{Kohn1959}%
  \BibitemOpen
  \bibfield  {author} {\bibinfo {author} {\bibfnamefont {W.}~\bibnamefont
  {Kohn}},\ }\href@noop {} {\bibfield  {journal} {\bibinfo  {journal} {Phys.
  Rev.}\ }\textbf {\bibinfo {volume} {115}},\ \bibinfo {pages} {809} (\bibinfo
  {year} {1959})}\BibitemShut {NoStop}%
\end{thebibliography}%

\end{document}